\documentstyle[multicol,aps,prl,epsf,graphics,psfig]{revtex}
\begin{document}
\draft
\preprint{}
\title{Critical Currents of Ideal Quantum Hall Superfluids}
\author{M. Abolfath$^{1}$, A. H. MacDonald$^{1}$, and Leo Radzihovsky$^2$}
\address{$^{1}$Department of Physics, University of Texas at Austin,
Austin TX 78712}
\address{$^{2}$Department of Physics, University of Colorado, Boulder,
  CO 80309}
\date{\today}
\maketitle
\begin{abstract}
  Filling factor $\nu=1$ bilayer electron systems in the quantum Hall regime 
  have an excitonic-condensate 
  superfluid ground state when the layer separation $d$ is
  less than a critical value $d_c$.  On a quantum Hall plateau current
  injected and removed through one of the two layers drives a
  dissipationless edge current that carries parallel currents, {\em
  and} a dissipationless bulk supercurrent that carries opposing
  currents in the two layers.  In this paper we discuss the theory of
  finite supercurrent bilayer states, both in the presence and in the
  absence of symmetry breaking inter-layer hybridization.  Solutions
  to the microscopic mean-field equations exist at all condensate
  phase winding rates for zero and sufficiently weak hybridization
  strengths.  We find, however, that collective instabilities occur
  when the supercurrent exceeds a critical value determined primarily
  by a competition between direct and exchange inter-layer
  Coulomb interactions.  The critical current is estimated using a local
  stability criterion and varies as $(d_c-d)^{1/2}$ when $d$
  approaches $d_c$ from below.  For large inter-layer hybridization,
  we find that the critical current is limited by a soliton
  instability of microscopic origin.
\end{abstract}

\begin{multicols}{2}
\columnwidth3.4in
\narrowtext
\section{Introduction}

In bilayer quantum Hall systems, broken symmetry ground
states\cite{halperin,rasoltprb} that have spontaneous interlayer phase
coherence were predicted some time ago
\cite{fertig89,yoshioka90,macdonald90,dlreview,wenzee1,ezawa}.  This
broken symmetry state is expected to be most robust near total Landau
level filling factor $\nu =1$ and occurs only when interactions
between electron in opposite layers are comparable in strength to
interactions between electrons in the same layer.  The putative existence 
of this broken symmetry was used several years ago to
explain\cite{yangmoon} the observation of a surprisingly strong
dependence\cite{murphy} of the bilayer system $\nu =1$ charged
excitation gap on in-plane magnetic field strength.

Spontaneous coherence between electrons in different energy bands is
an old topic in condensed matter physics, although it has not yet been
convincingly demonstrated outside of the quantum Hall regime.  For
example, it has long\cite{oldkeldysh} been realized that spontaneous
coherence at zero magnetic field is a possibility when overlapping or
nearby bands have opposite quasiparticle energy {\it vs.} wavevector
curvatures, the conduction and valence bands in a semiconductor or a
semimetal in particular.  Recent studies of optically generated
electron-hole plasmas in semiconductors, do indeed hint\cite{butov} at
the expected collective behavior.  In a separate materials system, the
discovery of weak ferromagnetism in lightly-doped divalent hexaborides
\cite{Young99}, which are ferromagnetic despite the absence of
partially filled {\it d}- or {\it f}- orbitals, led Zhitomirsky {\it
et al.} \cite{Zhitomirsky99} to propose recently that spontaneous
coherence between conduction and valence bands could be the mechanism
responsible for their ferromagnetism.  Spontaneous coherence between 
different hyperfine states in Bose-Einstein condensates of
magnetically confined $^{87}$Rb atoms, manifested by Rabi oscillations
between the two-components, is also closely related to the quantum Hall
bilayer phenomena studied here.\cite{spinorBEC}

Spontaneous coherence states are most commonly described using the
language of semiconductor physics in which a particle-hole
transformation is performed for the valence band; spontaneous
coherence between the bands then maps to electron-hole pair
condensation, something that is closely analogous to Cooper pair
condensation in a superconductor. These ordered states can be
equivalently described as a pseudo-spin-$1/2$ quantum ferromagnets
with the two states (top or bottom layer label in a quantum Hall
bilayer system, band-index in a bulk semiconductor, or alkali atom
species label in an ultra-cold atom system) defining an ordered
spinor\cite{yangmoon}.  The ordered state has superfluid properties
for ``staggered'' currents that flow in opposite directions in the two
bands, excitonic superfluidity in the language of semiconductors.  It
was suggested more than twenty years ago\cite{oldrussian} that
spontaneous coherence could occur between two-dimensional (2D)
conduction and valence bands localized in separate quantum wells and
that\cite{sfcond,paquetriceueda}, at least in mean-field theory, the
conditions required for condensation were more likely to be met when
the 2D systems experienced a strong perpendicular magnetic field.
It has not always been recognized, however, that because of the 
dispersionless Landau bands that occur in a strong
magnetic field, there is no difference\cite{yoshioka90} in this regime
between spontaneous coherence between a conduction band and a valence
bands and spontaneous coherence between two conduction bands.  The
spontaneous coherence that occurs in bilayer quantum Hall systems is,
in fact, precisely that originally anticipated by early theoretical
work.\cite{sfcond,paquetriceueda}.  Advances in our understanding of
quantum Hall physics have, however, given us a deeper appreciation of
the limitations of the mean-field theory approach used in the older
work and of the physics behind its partial success. Superflow of the
electron-hole pair condensate is perhaps the most characteristic
property of the bilayer quantum Hall broken symmetry states, which we
refer to here as quantum Hall superfluids.  In this paper we discuss
predictions for the maximum sustainable (counter-flowing) staggered
supercurrents supported by these states that follow from microscopic
mean-field theory for the ideal case in which the two-dimensional
electron layers are completely free from disorder.
\begin{figure}
\center
\epsfxsize 9cm \rotatebox{0}{\hspace{-0.3cm}\epsffile{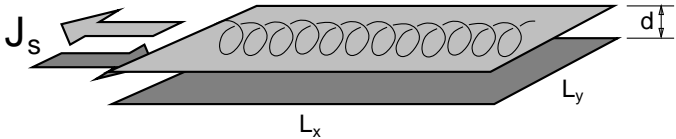}}
\vskip.5cm
\caption{A schematic illustration of a bilayer quantum Hall bar. For
  $\nu=1$ and sufficiently small interlayer spacing $d$ this system
  exhibits spontaneous interlayer phase coherence and excitonic
  superfluidity that supports counter-flowing (staggered) supercurrent
  $J_s$. The spiral is intended to indicate a uniform interlayer phase
  gradient that leads to constant $J_s$ along $x$.}
\label{bilayer}
\end{figure}
The work presented here complements an earlier field-theoretic
investigation by Kyriakidis and one us\cite{JordanLeo}, who studied
the decay of staggered supercurrents in bilayers quantum Hall
superfluids by thermally activated phase slips and calculated the
staggered I-V characteristics and critical current as function of
interlayer gate voltage. The main technical difference between the two
studies is that in the work of Ref.\onlinecite{JordanLeo} it was
assumed that the phase slip instability of a super-current carrying
state is at long wavelengths, an assumption that turns out to be valid
only for screened short-range interactions.  As we explicitly show
here, for realistic (long-range) Coulomb interactions, the instability
is at a finite wavevector. Furthermore, we argue that even with
short-range electron-electron interactions, when the interlayer
spacing $d$ is near the critical value $d_c$ (determined by the
transition out of interlayer coherent quantum Hall state), the
instability is also at short microscopic scales of order the magnetic
length.  This finding limits the range of validity of the field
theoretic approach of Ref.\onlinecite{JordanLeo} to the case of very
closely spaced layers, which cannot yet be realized experimentally.

Recently experimental advances by Eisenstein and
collaborators\cite{eisosaka,jpe1,jpe2,jpe3} have revealed several
dramatic effects that are only partially understood but are believed
to be due to collective transport effects in quantum Hall superfluids.
It appears, however, that the samples studied in current experiments 
have only short-range order because of disorder.  Their progress, which we
believe will open up a set of subtle new questions
about non-equilibrium properties of superfluid-like ordered
states, provides powerful motivation for the present work.  The key
capability which allows collective particle-hole transport effects to
be probed by electrical experiments is the possibility of making
separate electrical contact to two-dimensional electron gas layers
with a separation $d\approx 20 {\rm nm}$, small enough that is to be
in the range where spontaneous coherence occurs.  The strong zero-bias
peaks they see in interlayer tunneling conductance\cite{jpe1} studies
partially confirm predictions\cite{wenzee2,ezawa,tuntheory} of
Josephson-like effects, although these experiments are not fully
understood and two of us has argued elsewhere that the analogy with
the dc Josephson effect is incomplete.\cite{joglekar,ramintunneling}.
The evolution of the tunneling peak when the in-plane field\cite{jpe2}
is varied shows evidence of the
predicted\cite{fertig89,wenzee1,tuntheory,quinn} linearly dispersing
Goldstone collective mode associated with excitonic superfluidity.
Most directly related to the present theoretical paper, is a very
recent experiment\cite{jpe3}, which studies transport properties of a
quantum Hall superfluid for the case in which current (integrated
across the sample) flows through only one of the two layers.  Closely
related transport phenomena have also been discussed
theoretically\cite{bonsager} for the case of thin film ferromagnets.
\begin{figure}
\center
\vspace{-1.7cm}\epsfxsize 6.0cm \rotatebox{-90}{\epsffile{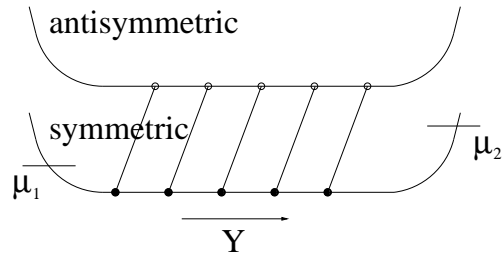}}
\caption{Schematic non-equilibrium quasiparticle populations for the
  quantum Hall effect in a Hall bar geometry for an ideal
  disorder-free quantum Hall bilayer superfluid.  The quasiparticle
  states are labeled by a guiding center coordinate which varies over
  the range from one edge of the sample to the other and is
  proportional to canonical momentum along the Hall bar.  The
  quasiparticle current is equally divided between the two layers and
  a net current is maintained by having different chemical potentials
  on the two edges.  The quasiparticles have spontaneous phase
  coherence and carry edge currents that are evenly partitioned
  between the two layers.  In order to satisfy the drag experiment
  constraint that no net current to flow in one of the layers, the
  condensate must carry an electron-hole supercurrent equal in
  magnitude to the quasiparticle current.  Unlike the charge current,
  the electron-hole supercurrent in the ideal case will flow uniformly
  through the bulk of the system.  The mean-field state of a quantum
  Hall superfluid pairs electrons and holes that have different
  momenta in the current carrying direction, or equivalently,
  different cyclotron-motion guiding centers. This property is
  indicated schematically by the slanted lines that connect different
  guiding centers.}
\label{fig0}
\end{figure}
The quantum Hall effect and superfluidity share the unusual property of
transport without dissipation.  In superfluids, dissipationless transport is
possible because the quasiparticles are in equilibrium with a current-carrying
condensate. In the quantum Hall effect, as illustrated schematically in
Fig.~\ref{fig0}, the quasiparticles are not in equilibrium.  Instead the gap
for charged excitations in the bulk implies that low-energy quasiparticles are
localized at the sample edges, allowing a net current to be carried through
the system \cite{Halperin82} without dissipation by maintaining a Hall voltage
difference between isolated subsystems on opposite edges.  In the mean-field
theory\cite{matsuepaper} of a quantum Hall superfluid, the occupied
quasiparticle state wavefunctions are coherent linear combinations of orbitals
localized in separate layers.  For equal density in the bilayers, the current
they carry is divided equally between the layers and any voltage probe that
couples to the quasiparticle system will measure the same value in either
layer, leading to large drag voltages\cite{vignaledragpaper}.  Since the
quasiparticles carry equal current in the two layers, the only way in which it
is possible to have no net current in one of the layers, is to have it
canceled by spontaneously generated staggered supercurrent, as illustrated in
Fig.~\ref{dragfig}. In fact, as discussed in Ref.\onlinecite{kunyang,girvin},
since both staggered supercurrent and the uniform Hall current are
dissipationless, at $\nu=1$ an ideal quantum Hall superfluid state should
exhibit, respectively, vanishing and quantized longitudinal and Hall
drag-resistivities.  

\begin{figure} 
\center 
\vspace{-.5cm} 
\epsfxsize 9cm \rotatebox{0}{\hspace{-0.3cm}\epsffile{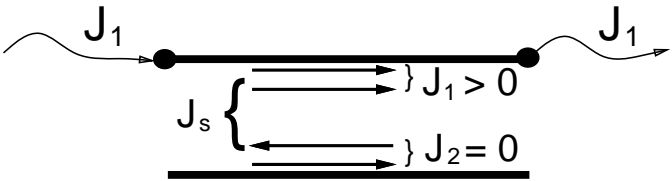}}
\vspace{.5cm} 
\caption{Schematic for an integer quantum Hall superfluid bilayer in the 
  drag geometry of Kellogg, et al., Ref.23, where electrical current
  $J_1$ is only allowed to flow in the top layer.  To satisfy the
  boundary condition of vanishing current in the bottom layer,
  $J_2=0$, the system spontaneously develops a dissipationless
   staggered current $J_s$.}
\label{dragfig} 
\end{figure}

Current experiments are not fully in accord with this simple picture;
the voltage drops in the current carrying direction are not zero and
not identical in the two layers, and the Hall voltages are not
perfectly quantized as one would expect in a $\nu=1$ quantum Hall
bilayer superfluid at $T=0$.  Most likely the discrepancy is due to
disorder which leads\cite{ourcomingpaper} to mobile quasiparticle
excitations in the bulk of the two-dimensional electron system and to
vortex flow and consequent phase slips in the superfluid order
parameter field.  Long length scale inhomogeneity in the samples that
leads to phase separation between ordered and disordered regions can
also be invoked to explain\cite{halperinstern02} many aspects of
current experimental results.  The calculations we report in this
paper do not account for disorder and do not attempt to explain
quantitative aspects of current experiments.  Instead we focus on the
properties of ideal systems in which no disorder is present, with the
expectation that the ideal situation will be approached more closely
in the future.  We expect the collective instability
studied here to control longitudinal dissipation and deviation from
Hall-drag quantization in clean quantum Hall bilayer superfluids.

In Section II of this paper we briefly summarize the mean-field-theory
of current-carrying states in quantum Hall superfluids, allowing for
the possibility of explicit symmetry breaking inter-layer
hybridization terms in the microscopic Hamiltonian and for weak links
in the two-dimensional superfluid that are created by gate voltages.
We define the critical current as the maximum current at which stable
local minima in the energy-functional of the quantum Hall superfluid
exist; formal expressions for the stability matrix at mean-field
energy-functional extrema are also given in Section II.  Some of these
formal mean-field-theory results have been discussed in another
context in earlier work, \cite{RaminLeo} but are included here for
completeness.  In Section III we apply these formal results to the
case of a uniform quantum Hall superfluid with no inter-layer
hybridization.  In this case we are able to derive physically
transparent expressions for the critical current.  An important
prediction that follows from this analysis is the way the critical
current vanishes as the phase boundary that separates the quantum Hall
superfluid from normal ground states of the 2D electron systems is
approached.  Section IV deals with the potentially interesting
case\cite{wenzeeepl} in which weak links are intentionally induced in
these two-dimensional electron systems by applying voltages to
external gates.  We find that even relatively weak disturbances lead
to drastically reduced critical currents. This partially explains the
sensitivity of experimental samples to disorder, that will undoubtedly
introduce such weak links in the superfluid bilayer.  In this case we
are able to derive a Josephson-like relationship between the order
parameter phase jump across the weak link and the staggered
supercurrent that flows across it.

An important difference between
excitonic superfluids and both superconductors and fermion-pair
superfluids is the inevitable presence in the Hamiltonian of explicit
staggered gauge-symmetry breaking terms that allow electrons to tunnel
between layers and hybridize quasiparticle states even when there are
no interactions.  These terms can be extremely weak but are never
strictly zero.  In Section V we show that staggered current causes 
a soliton lattice to form at a characteristic current density. 
For very strong tunneling, we find that the maximum
supercurrent drops to zero.  We conclude in Section VI with a summary
of our results and a discussion of the role of quenched disorder and
thermally activated vortices and phase slips that is informed by our
numerical results for the microscopic properties of ideal systems.  We
conclude that non-linear transport in the {\em drag experiment}
geometry used by Eisenstein and collaborators\cite{jpe3}, studied as a function
of carrier density, might be able to identify a crossover in dominant
dissipation mechanism that is closely associated with the ideal
system critical currents that are estimated in this paper.


\section{The Superfluid State Energy Functional: Mean-Field Equations
  and Stability Analysis}

All of the calculations summarized in this article are based on a
microscopic Hartree-Fock approximation (HFA) for the energy functional
of quantum Hall staggered superfluid states and are analogous to ones
made using the BCS theory for superconductors.  We will consider only
situations in which the order parameter is a function of a single
($X$) spatial coordinate, making it convenient to choose a Landau
gauge in which the Landau level orbitals are localized in this
direction.  Because we ignore edge effects, it will prove more
convenient to choose orbitals localized along the direction of current
(X) flow rather than the more conventional choice perpendicular to the
direction of current flow illustrated in Fig.\ref{fig0}.  The energy
functional is $E_{MF}[\theta, \varphi] =\langle \Psi| \hat{T} +
V_{ext} + V_{ee}|\Psi\rangle$ where $T$ is the interlayer tunneling
term, $V_{ext}$ is the external potential term, and $V_{ee}$ is the
Coulomb interaction in the microscopic Hamiltonian.  The variational
wavefunction used to define this energy functional is
\begin{equation}
|\Psi\rangle =\prod_X \left(\cos[\theta(X)/2] \hat{c}^\dagger_{X T} +
  \sin[\theta(X)/2] e^{i\varphi(X)} \hat{c}^\dagger_{X B}
  \right)|0\rangle,
\label{PsiHF}
\end{equation}
where $X$ is a Landau gauge guiding center label, $T$ (top) and $B$
(bottom) are layer indices, and $\theta$ and $\varphi$ define the
coherence factors of the BCS-like single-Slater-determinant
many-particle wavefunction.  If the phase coherent state is regarded
as a pseudospin ferromagnet, the angles $\theta(X)$ and $\varphi(X)$
are polar and azimuthal angles that specify the direction of the
pseudospin magnetization at position $X$.  This is a special case of a
class of variational wavefunctions in which virtual charge density
fluctuations are not permitted. The energy functional can be written
explicitly\cite{Cote94,Kyriakidis,RaminLeo} in terms of Hartree and
exchange microscopic two-particle interaction matrix elements:
\end{multicols}

\widetext

\begin{eqnarray}
E_{MF} = 
-\frac{\Delta_{SAS}}{2} \sum_{X} \cos\varphi(X) && \sin\theta(X)
- \sum_{X} V_g(X) \cos\theta(X)
+ \frac{1}{2L_y}\sum_{X,X'}\left[H(X-X')-\frac{1}{2}F_S(X-X')\right]
\cos\theta(X) \cos\theta(X')
\nonumber \\&&
-\frac{1}{4L_y}\sum_{X,X'}F_D(X-X') \sin\theta(X) \sin\theta(X')
\cos[\varphi(X)-\varphi(X')],
\label{HF}
\end{eqnarray}
\noindent
where $\Delta_{SAS}$ is the splitting between symmetric and
antisymmetric orbitals induced by hybridization, $V_g(X)$ is the
external gate voltage averaged over guiding center orbital $X$, and
$H,~F_S$, and $F_D$ are the Hartree matrix element, the same-layer
exchange matrix element, and the different-layer exchange matrix
element.  If we neglect the finite widths of the quantum wells the
interaction matrix elements are given by:
\begin{mathletters} 
\begin{eqnarray} 
H(X)&=&\int \frac{dq}{2\pi} \frac{2 \pi e^2 (1 - e^{-qd})}{2\epsilon q} 
e^{iqX} e^{-q^2\ell^2/2},\label{hofX}\\  
F_C(X)&=&e^{-X^2/2\ell^2} \int\frac{dq}{2\pi} V_C(q,X/\ell^2) 
e^{-q^2\ell^2/2}, 
\label{fofx} 
\end{eqnarray} 
\end{mathletters} 

\begin{multicols}{2}
\columnwidth3.4in
\narrowtext
\noindent
where $V_C(q_x,q_y) = 2 \pi e^2/(\epsilon q)$ and $2\pi e^2
\exp(-qd)/(\epsilon q)$ for $C=S$ and $C=D$ respectively. Note that
the exchange integral drops rapidly with orbit center separation,
while the electrostatic integral falls only as $X^{-2}$ at large $X$,
corresponding to the interaction between dipole moment lines created
by charge transfer between the layers.  In these equations $\ell
\equiv \sqrt{\hbar c/eB}$ is the magnetic length.  The Fourier
transform of these functions will figure prominently in analyzing the
superfluid properties of uniform bilayers. The energy functional
$E_{MF}[\theta(X),\varphi(X)]$ can be elevated to a quantum theory by
adding to it a Berry phase contribution to the
Lagrangian\cite{yangmoon,burkov}.

We choose the convention
\begin{equation}
H(p) = \int {d X\over 2 \pi\ell^2} \exp(i p X) H(X)
\end{equation}
(and correspondingly for the exchange integrals) so that these
quantities have units of energy.  When finite-thickness effects are
neglected, we obtain the following explicit expressions for the
Coulomb interaction case:

\begin{mathletters}
\begin{eqnarray}
H(p)&=&\frac{e^2}{2\epsilon \ell^2} e^{-p^2\ell^2/2} \frac{1-e^{-d|p|}}{|p|},\\
F_D(p)&=&\frac{e^2}{\epsilon} \int_0^\infty dr J_0(rp/\ell^2) 
e^{-r^2 \ell^2/2} e^{-rd},
\end{eqnarray}
\end{mathletters}
where $J_0(x)$ is the Bessel function. The same layer exchange
function $F_S(p)$ differs only by the absence of the $e^{-rd}$ factor.
$F_D(p)$ is plotted in Fig. \ref{Js0}.

\subsection{Mean-field Equations}
Mean-field states of quantum Hall superfluids are extrema of the
energy functional (\ref{HF}) and satisfy mean-field equations obtained
by setting $\delta E_{MF}/\delta\varphi(X)=0$, and $\delta
E_{MF}/\delta\theta(X)=0$.  These Euler-Lagrange equations for the
variables $\theta$ and $\varphi$ are coupled, and highly non-linear.
The equations are, in addition, strongly non-local because of the long
range of the Coulomb interaction term.  For $V_{g}(X)$ and
$\Delta_{SAS}$ equal to zero, the global minimum of the energy
functional is $\theta(X) = \pi/2$ and $\varphi(X)$ equal to an
arbitrary constant.  Since we will be interested primarily in
solutions with $\theta(X)$ near $\pi/2$, it is useful to define
$\eta(X)=\pi/2 - \theta(X)$.  For our investigation we will be
interested only in extrema that are periodic functions of $X$.
Denoting the period by $a$ and using the thermodynamic limit property
$\sum_X \rightarrow L_y \int dX/(2\pi\ell^2)$, the mean-field
equations can be written as:
\end{multicols}
\widetext

\begin{mathletters}
\label{CHF}
\begin{eqnarray}
\tan\varphi(X)&=&
\frac{\int_0^a\frac{dX'}{2\pi\ell^2}\cos\eta(X')\sin\varphi(X')
\sum_{n=-\infty}^\infty 
F_D(X-X'-na)}{\Delta_{SAS}+\int_0^a\frac{dX'}{2\pi\ell^2}
\cos\eta(X')\cos\varphi(X')\sum_{n=-\infty}^\infty F_D(X-X'-na)},\\
\label{CHF1}
\tan\eta(X)&=&\frac{2V_g(X)+\int_0^a\frac{dX'}{2\pi\ell^2}\sin\eta(X')
\sum_{n=-\infty}^\infty[F_S(X-X'-na)-2H(X-X'-na)]}
{\Delta_{SAS}\cos\varphi(X)+\int_0^a\frac{dX'}{2\pi\ell^2}
\cos\eta(X')\cos[\varphi(X)-\varphi(X')]\sum_{n=-\infty}^\infty F_D(X-X'-na)}.
\label{CHF2}
\end{eqnarray}
\end{mathletters}

\begin{multicols}{2}
\columnwidth3.4in
\narrowtext
\noindent

Our primary interest here is in current carrying states, and we choose
$a$ to be the distance over which the condensate phase increases by $2
\pi$; our mean-field states are ones in which phase slips form a
lattice.  Note that $\sin (\varphi(X))$, $\cos (\varphi(X))$ and $\eta
(X)$ are periodic.  It follows that the total phase change of the
order parameter on going from one end of the system to the other is $2
\pi L/a \equiv 2 \pi N_{w}$

These equations must be solved self-consistently to locate mean-field
states.  They can be understood most simply in the language of
pseudo-ferromagnetism, in which they simply state that at each
position $X$, the orientation of the quasiparticle pseudo-spinor is
along the direction of the total pseudo-spin effective field,
including direct and exchange mean-field interaction and external
potential contributions felt by the quasiparticles at that point.
The quasiparticles of the superfluid are in equilibrium with the
condensate at extrema of the mean-field energy functional.

\subsection{General Expression for the Condensate Current}

The expression we use for the condensate staggered supercurrent
follows from arguments presented in greater detail in previous work on
quantum Hall superfluids\cite{yangmoon} and implicitly in work on
excitonic superfluidity.  The current operator in each layer can be
expressed in terms of the derivative of the Hamiltonian with respect
to the vector potential in that layer.  The vector potentials in the
two layers can be varied by independent gauge transformations for the
two layers and physical properties can depend on the phase difference
between the two layers only through the gauge-invariant quantity
$\hbar\nabla \varphi + (e/c) (\vec A_{T}- \vec A_{B})$ where $\vec
A_{T}$ and $\vec A_{B}$ are the vector potentials in top and bottom
layers and $\varphi(X)$ is the difference in wavefunction phase
between top and bottom layers that appears in the HF variational
wavefunction, Eq.\ref{PsiHF}.  Since the operator for the spatially
averaged current in a layer is proportional to the derivative of the
Hamiltonian with respect to the vector potential in that layer and we
have chosen the same gauge for the vector potential for each layer, it
follows that the spatially averaged condensate current is proportional
to the rate of change of condensate energy with $N_w$:
\begin{equation}
J_s = \frac{1}{\pi \hbar} \frac{d E_{MF}}{L_y d N_{w}} = 
-\frac{a^2}{\pi \hbar} \frac{d \varepsilon_{MF}}{d a}, 
\label{eq:current2}
\end{equation}
where $\varepsilon_{MF}$ is the energy per cross-sectional area.  This is
the expression that we use below to extract condensate staggered
currents from solutions of the mean-field equations.  In
Eq.~\ref{eq:current2} $J_s$ is a number staggered current density,
{\em i.e.} it corresponds to an electrical current density equal to
$-e J_s(X)$ in the top layer and $e J_s(X)$ in the bottom layer.  When
tunneling between the layers is allowed, there is a circulating
current within each period of the phase slip lattice in addition to
this spatially averaged current.  The circulating current is
conveniently calculated from charge conservation which relates the
current flowing between the layers to the divergence of the condensate
supercurrent.  The experimental meaning of the condensate current 
evaluated in this way is discussed at greater length in Section VI.

\subsection{Stability Analysis} 

We restrict ourselves to the case of zero bias voltage
for which the mean field solutions will have $\eta(X) \equiv 0$.
A mean-field solution is a local {\em minimum} of the energy
functional provided that the stability matrices
\begin{mathletters}
\begin{equation}
K_{\varphi\varphi}(X, X') =
\frac{\delta^2 E_{MF}}{\delta\varphi(X)\delta\varphi(X')},
\label{Kff}
\end{equation}
and 
\begin{equation}
K_{zz}(X, X') = \frac{\delta^2 E_{MF}}{\delta\eta(X)\delta\eta(X')},
\label{Kzz}
\end{equation}
\end{mathletters}
are both positive definite.  The explicit expressions for the
stability matrices evaluated at the mean field solutions $\eta(X) \equiv 0$,
and $\varphi=\varphi(X)$ are as follows:
\end{multicols}
\widetext

\begin{eqnarray}
K_{zz}(X,X') &=& \frac{1}{2 L_y} \left[2H(X-X')-F_S(X-X')\right]
\nonumber \\ 
&& + \frac{1}{2}\delta_{X,X'} \left[\Delta_{SAS} \cos(\varphi(X))
+ {1\over L_y} \sum_{X''} F_D(X-X'') \cos(\varphi(X)-\varphi(X''))\right].
\label{Kzzexpression}
\end{eqnarray}
\begin{eqnarray}
K_{\varphi\varphi}(X,X') &=& -\frac{1}{2 L_y} F_D(X-X')
\cos(\varphi(X)-\varphi(X')) \nonumber \\ &&
+ \frac{1}{2}\delta_{X,X'} \left[\Delta_{SAS} \cos(\varphi(X))
+ {1\over L_y} \sum_{X''} F_D(X-X'') \cos(\varphi(X)-\varphi(X''))\right].
\label{Kffexpression}
\end{eqnarray}

\begin{multicols}{2}
\columnwidth3.4in
\narrowtext
\noindent
 
As we show in the next section, for $\Delta_{SAS}=0$, the mean-field
solutions do not break translational invariance, and the stability
matrices can be diagonalized explicitly by taking advantage of the
translational symmetry property.  The spectrum of the stability
matrices can then be related to the Fourier transformed direct and
exchange interaction matrix elements as we show below.  In the more
general case of finite $\Delta_{SAS}$, it is necessary to evaluate
their spectrum numerically, which we do in Sec.\ref{finiteDelta} by
discretizing the set of allowed guiding centers.

\section{Critical Current of Uniform Quantum Hall Superfluids 
($\Delta_{SAS}=0$)}

\begin{figure}
\center
\vspace{-1.cm}
\epsfxsize 6.0cm \rotatebox{-90}{\epsffile{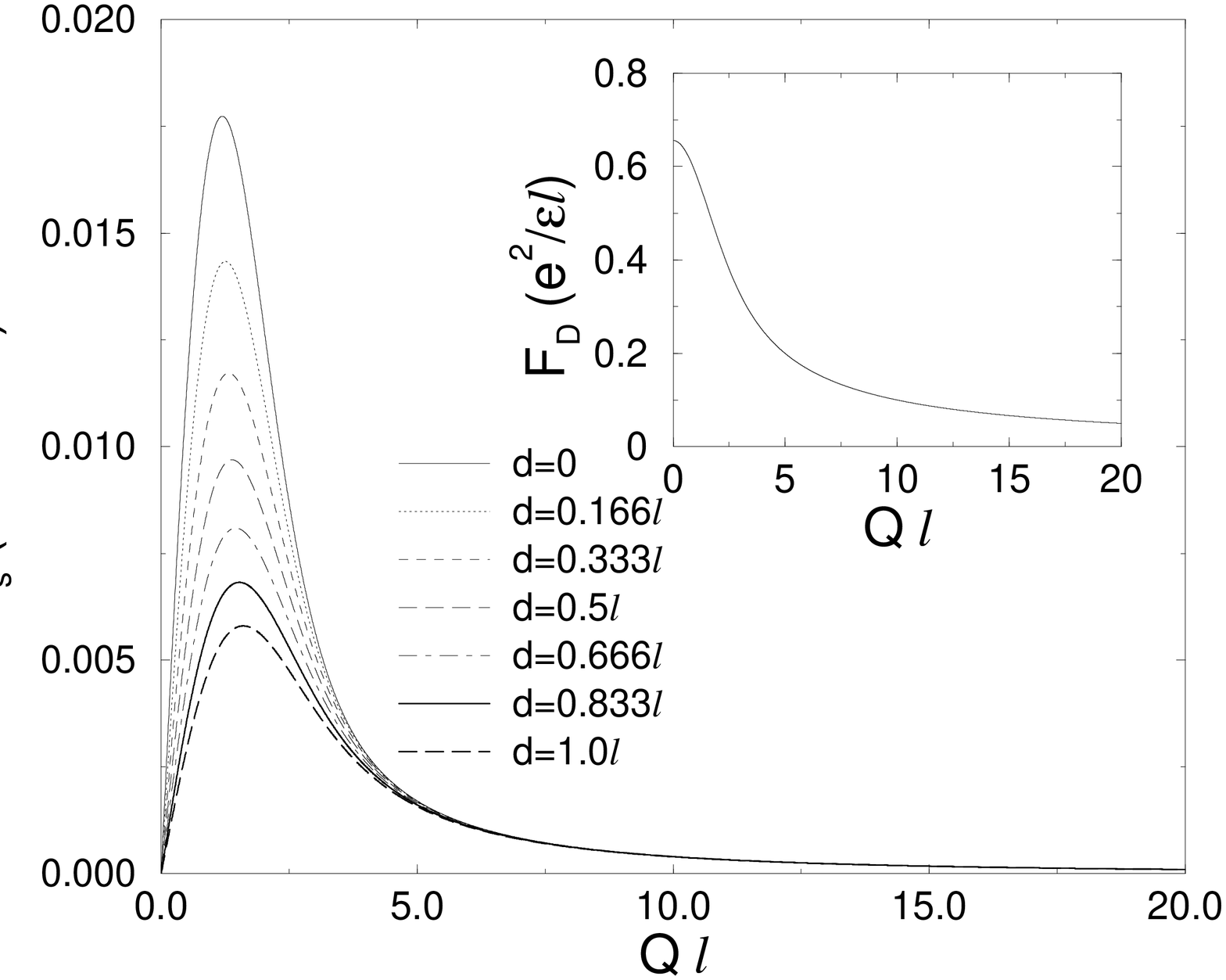}}
\setlength{\unitlength}{1mm}
\begin{picture}(150,25)(0,0)
\put(20,33) {$|$}
\put(18,29) {$Q^*l$}
\end{picture}
\vspace{-2.6cm}
\caption{$J_s$ vs. pairing wavevector $Q$ for $\Delta_{SAS}=V_g=0$ and a 
  series of layer separation $d$ values.  $J_s$ is in units of $J_0/e
  \equiv \frac{1}{\hbar} \frac{e^2}{\epsilon \ell^2}$ ($J_0 = 41
  \times 10^{-6}{\rm Amp}/\mu m B[{\rm Tesla}]$.)  Our stability
  analysis demonstrates that the mean-field state is always unstable
  for pairing wavevectors larger than $Q^*$, the pairing wavevector at
  which $J_s$ is maximized.  For quantum Hall superfluids, however,
  the critical current is almost always limited by unstable amplitude
  fluctuations of the electron-hole pair condensate that are indicated 
  by negative values of $K_{zz}(q)$.
  For example for $d = \ell$, the maximum supercurrent mean-field
  state ($J_{c} = 0.0058 J_0$) occurs at $Q\ell = 1.6$, but amplitude
  fluctuations are unstable for $Q \ell > 1.2$ as we discuss below.
  The inset shows $F_D(Q)$ vs. $Q$ for $d = \ell$.  $F_D(Q) \sim
  constant -Q^2$ at small $Q$ and $ \sim 1/Q$ at large $Q$ so that
  $J_s$ is proportional to $Q$ and $Q^{-2}$ at large and small $Q$
  respectively.}
\label{Js0}
\end{figure}

For $\Delta_{SAS}=0$ and $V_{g}(X)$ equal to a constant, there are
solutions to the mean-field equations for which $\eta$ (bilayer charge
imbalance) is spatially uniform and $\varphi(X)$ varies at a constant
rate.  The largest possible supercurrents flow for balanced bilayers,
{\it i.e.} for $V_{g}(X) \equiv 0$.  The family of solutions with
$\eta=0$ and $\varphi=Q X$ corresponds to states in which electron-hole
pairs have condensed into a state with total momentum $Q \hat x$.
Unlike the case of the zero-field fermion pair condensates, solutions
to the mean field equations can be found for any pairing momentum $Q$,
no matter how large.  Inserting these solutions in Eq.(\ref{HF}), we
find that the condensate-dependent part of the energy per unit area
for pairing at wavevector $Q$ is $F_D(Q)/ 8 \pi \ell^2$.  The
condensate current then follows from Eq.(\ref{eq:current2})
\begin{equation}
J_s = \frac{2}{\hbar} \frac{d\varepsilon_{MF}}{dQ}=\frac{-1}{4\pi\hbar\ell^2}
\frac{2 d F_D(Q)}{dQ}
\label{eq5_1}
\end{equation}
The dependence of $J_s$ on pairing wavevector $Q$ is illustrated in
Fig. \ref{Js0} for a range of bias voltages.  The natural unit of
charge current density in the quantum Hall regime is $J_0 \equiv
2 e^3/\hbar \epsilon \ell^2 \approx 82 {\rm \mu A/ \mu m} B [\rm
Tesla]$.  
The main effect of a uniform bias voltage is to reduce the
scale for supercurrent values.  We should also remark that these
supercurrent values and all numerical results discussed in this paper
will be influenced somewhat by corrections that account for the finite
thickness of the two-dimensional layers.  These corrections could be
incorporated into our calculations without any difficulty, but we
choose to ignore them here mainly for the sake of simplicity.  As we
discuss later, other more fundamental and difficult issues arise when
we attempt to compare this analysis with experiments in real systems.

The Hamiltonian $H_f$ for Gaussian fluctuations $\phi(X)=\varphi-Q X$
and $m_z(X)$ around these uniform superfluid ($\varphi_0=QX$, $m_z=0$)
mean-field states can be expressed in terms of the stability matrices
discussed previously.  In the uniform current case, $K_{\varphi
\varphi}(X,X')$ and $K_{zz}(X,X')$ depend only on $X-X'$ and the
matrices can be diagonalized by Fourier transformations.  

We find that
\begin{eqnarray}
\frac{H_f[\varphi, m_z]}{L_x L_y} &=& \frac{1}{2} \sum_q \varphi(-q)
K_{\varphi\varphi}(q) \varphi(q) \nonumber \\ 
&& + \frac{1}{2}\sum_q m_z(-q) K_{zz}(q) m_z(q),
\label{fluc}
\end{eqnarray}
\noindent
where
\begin{mathletters}
\begin{eqnarray}
K_{\varphi\varphi}(q)&=&\frac{1}{2\pi \ell^2}\left[
F_D(Q) -\frac{F_D(q+Q)+F_D(q-Q)}{2}\right],\\
\label{eq3n}
K_{zz}(q)&=& \frac{1}{2\pi\ell^2}\left[H(q) - 
\frac{1}{2}F_S(q) + \frac{1}{2}F_D(Q)\right].
\label{eq4n}
\end{eqnarray}
\end{mathletters}

\begin{figure}
\center
\epsfxsize 6.0cm \rotatebox{-90}{\epsffile{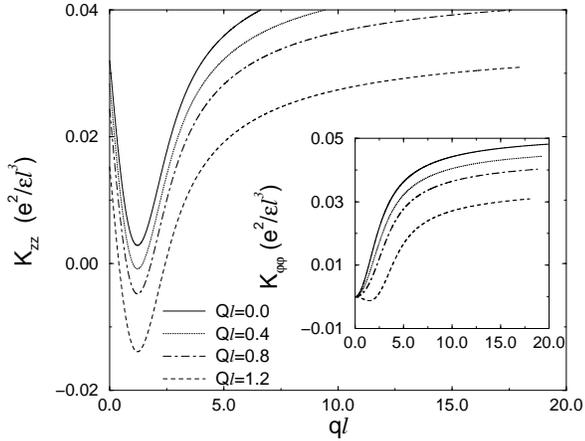}}
\vskip.5cm
\caption{
  $K_{zz}(q)$ and $K_{\varphi\varphi}(q)$ vs. $q$ for $d=\ell$, and
  $\Delta_{SAS}=0$.  $K_{\varphi\varphi}(q=0)=0$ even at finite
  pairing momentum $Q \ne 0$, because the energy is invariant under a
  spatially constant change in the phase $\varphi$.  $K_{zz}(q)$ is
  nonanalytic at $q=0$, decreasing in proportion to $|q|d$, because of
  long-range Coulomb interactions which favor non-uniform exciton
  density variations.  The decrease of $K_{zz}(q)$ for $q d$ small is
  rapid for large layer separations $d$, and eventually drives the
  critical current to zero, destabilizing the superfluid state.  }
\label{Kzz_a20}
\end{figure}
(In these equations we have used the following conventions for the
Fourier transforms of $\phi(X)$ and $m_z(X)$.  $\varphi(q) \equiv
\sum_X \phi(X) exp(-i p X)/ N_{\phi}$ where $N_{\phi} = L_x L_y / (2
\pi \ell^2)$ is the total number of Landau gauge orbitals in the
system.  This choice is made so that $\varphi(q)$ is dimensionless and
$\varphi(q=0)$ is the spatial average of $\phi(X)$.)  The first two
terms in Eq.(\ref{eq4n}) represent the cost in electrostatic energy
and the gain in exchange energy that accompanies fluctuations in the
bilayer charge balance.  Both stability kernels $K_{\varphi\varphi}$
and $K_{zz}$ must be positive definite for the superfluid state to be
stable against small fluctuations.  From Eq.(\ref{eq3n}), we see that for
$q \to 0$, $K_{\varphi\varphi} \to 4 \hbar q^2 \frac{\partial
  J_s}{\partial Q}$, which is positive only on the increasing portion
of the $J_s(Q)$ curve; at larger pairing wavevectors phase separation
into regimes with larger and smaller average phase winding rates is
always energetically preferred over the homogeneous state.

\begin{figure}
\center
\epsfxsize 6.0cm \rotatebox{-90}{\epsffile{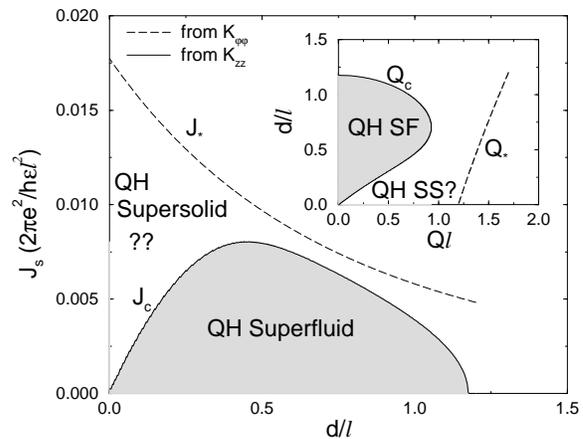}}
\vskip.5cm
\begin{picture}(150,25)(0,0)
\end{picture}
\vspace{-1.cm}
\caption{Critical supercurrent density $J_c(d)$ (main panel) and
  (inset) pairing wavevector $Q_c(d)$ of the critical current {\it
  vs.} layer separation $d$, defining the QH Superfluid phase boundary
  in pairing wavevector-layer separation space.  The dashed lines in
  the main figure and the inset showing critical current, $J_*$ and
  maximum pairing wavevectors, $Q_*$ obtained by applying the
  $K_{\varphi\varphi}$ (condensate phase) order parameter stability
  criterion is shown by these results to be less stringent; the solid
  line is determined by the $K_{zz}$ stability criterion.  For $d
  \approx d_c$, $J_c \propto (d_c-d)^{1/2}$, while for $d \to 0$
  $J_c \propto d$.  Current experimental samples have layer
  separations close to the critical value $d_c$. Because the
  $K_{zz}(q)$ instability is at a finite wavevector $q_*$, we expect
  that for small $d$ the resulting current carrying state is a {\em
  modulated} QH superfluid, i.e., a QH supersolid.  For larger values
  of $d$, the instability is at a large finite wavevector and we
  expect a first order transition between uniform superfluid and
  disordered ground states.}
\label{unisupercurrent}
\end{figure}

For $q \to 0$ and $Q=0$, $K_{\varphi\varphi} \to \rho_s q^2 $ where
$\rho_s$ is phase stiffness (superfluid density) parameter of the
superfluid.  $K_{\varphi\varphi}(q)$ appears to always be positive
when it is positive for $q \to 0$; the $K_{\varphi\varphi}$ stability
criterion $Q < Q^*$ (with $Q^*$ the pairing wavevector at which
$J_s(Q)$ is maximum) does not limit the critical current of quantum
Hall superfluid to values below the maximum value at $Q^*$ that occurs
in mean-field solutions.  In contrast, $K_{zz}(q)$ always has its
minimum at a finite value of $q$.  This differs from the
field-theoretic treatment of Ref.\onlinecite{JordanLeo}, where
$K_{zz}(q)$ was taken to be a monotonically increasing function of
$q$.  This assumption is not valid for unscreened (long-range) Coulomb
interactions. The last term in the expression for $K_{zz}(q)$, which
represents the loss in condensation energy when the condensate
electron-hole density is varied from its optimal value, is essential
for the stability of quantum Hall superfluids at typical values of
$d/\ell$. As illustrated in the inset of Fig.\ref{Js0}, this
stabilizing term decreases in magnitude at finite $Q$ because of the
reduced condensation energy associated with finite $Q$ pairing.  It is
this reduction in condensation energy that limits the critical
staggered supercurrent in our theory.  It is also worth remarking that
$K_{zz}(q)$ is negative for any finite value of $Q$ for $d = 0$.  It
is only at $d = 0$ that the ground state is given exactly by the
Halperin\cite{halperin} $(1,1,1)$ two-component quantum Hall fluid
wavefunction.  It follows that the $(1,1,1)$ state is {\em not} a
superfluid, a point that we have made\cite{zhang} previously.  Because
of this property, field-theoretical analyses of quantum Hall
superfluid properties that start from $(1,1,1)$ variational
wavefunctions should be regarded with some caution, in our view.

In Fig.\ref{unisupercurrent} we have plotted the layer separation
dependence of the ideal bilayer critical current $J_c$ that follows
from this stability analysis.  The critical current vanishes as $d \to
0$ because of the vanishing difference between intralayer and
interlayer interactions that is necessary to provide a barrier to
phase slip nucleation.  In the language of
pseudo-ferromagnetism\cite{yangmoon,LRcanting}, the corresponding
observation is that easy-plane anisotropy is required for spiral state
metastability.  The staggered critical supercurrent also vanishes at
large layer separation because the stiffness against electron-hole
(bilayer charge imbalance) density waves vanishes as we discuss below.
This figure shows that the critical current is always limited by the
$K_{zz}(q)$ stability criterion.  The maximum critical current occurs
for $d \sim 0.4 \ell$ and is $\sim 0.008$ of our current-density unit
$J_0$.  It follows from this analysis that under typical experimental
circumstances the critical current of an ideal system should be $\sim
0.33 {\rm \mu A {\mu m}^{-1}}$. This staggered critical current value
is several orders of magnitude larger than the value used in recent
drag experiments\cite{jpe3}.
Although existing drag experiments provide evidence of collective
transport, they explicitly demonstrate dissipation in the pseudospin
current channel because the measured longitudinal electric fields in
the two layers are different.  This dissipation is likely due to the
flow of vortices induced by thermal fluctuations and disorder that we
do not account for here.  It would nevertheless be interesting to
experimentally explore the possibility that a large increase in the
difference in longitudinal electric fields would occur relatively
sharply at larger current densities, reflecting a change in the
dominant dissipation mechanism when the ideal critical current
addressed in this paper is exceeded.

The $K_{\varphi\varphi}(q)$, and $K_{zz}(q)$ stiffnesses in the
fluctuation Hamiltonian are inversely related to static response
functions.  Since the phase of the condensate and the electron-hole
pair density are canonically conjugate coordinates these two
stiffnesses are also simply related to the system's quantized
collective excitation energies.  The semiclassical picture of these
modes emerges from the Landau-Lifshits equations of motion of the
pseudospin ferromagnet, which can be derived by adding a Berry phase
term to the above fluctuation Hamiltonian to obtain the system's
Lagrangian\cite{yangmoon}.

\begin{figure}
\center
\epsfxsize 6.0cm \rotatebox{-90}{\epsffile{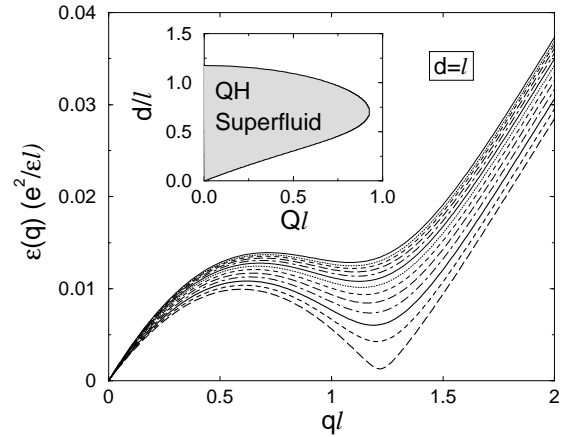}}
\vskip.5cm
\caption{The dispersion relation of the quantum Hall superfluid
  collective modes is plotted for $d=\ell$, and $V_g=\Delta_{SAS}=0$
  for values of the pairing wavevector $Q$ which vary between $Q=0$
  (the top curve) and $Q=0.7\ell^{-1}$.  At $Q=Q_c$, where the roton
  minima vanishes, the superfluid mean-field state is unstable.}
\label{omega_Q}
\end{figure}

The superfluid collective modes have energy $\varepsilon(q) = 2 \pi
\ell^2 \sqrt{K_{zz}(q) K_{\varphi\varphi}(q)}$, generalizing earlier
results\cite{paquetriceueda,fertig89,quinn,yangmoon} to the case of a
finite supercurrent state.  For all pairing wavevectors, the
collective modes have a linear dispersion of a superfluid at long
wavelengths.  At shorter length scales (large $q$) $\varepsilon(q)$ also
has a minimum, that is reminiscent of the roton minimum in superfluid
Helium.  The source of this characteristic minimum in the collective
mode dispersion $\varepsilon(q)$ is the Coulomb interaction, which leads
to the minimum in $K_{zz}(q)$ displayed in Fig.\ref{Kzz_a20}.  The
collective mode dispersion is altered when a supercurrent is flowing
primarily because of the reduction in the condensation energy and
$K_{zz}(q)$ with increasing $Q$.  When the minimum is already close to
zero for $Q=0$, the critical supercurrent is small.  Since $F_D(Q)$ is
a quadratic function of $Q$ for small $Q$ and the minimum in
$K_{zz}(q,Q=0)$ approaches zero linearly as $d \to d_{c}$, it follows
that $Q_c$ and the critical staggered supercurrent vanish like
$(d_{c}-d)^{1/2}$ for $d \to d_{c}$.

The physics that limits the ideal critical current of quantum Hall
superfluids is similar to that encapsulated by Landau's expression for
the critical current of superfluid Helium, in which he argued that the
superfluid velocity cannot exceed $J_s^{max} = v_s^{max} n =
\min{(\varepsilon^*(q)/\hbar q)} n$, where $\varepsilon^*(q)$ is the
elementary excitation dispersion relation when no supercurrent is
flowing, {\em i.e.} at $Q=0$, and the minimum above is over all
wavevectors $q$.  In both superfluid Helium and quantum Hall superfluid 
cases, Landau's argument approximates the more fundamental
requirement that the order parameter of the current-carrying state be
at a local minimum of the model's energy functional.  The remark that
we make here applies {\em mutatis mutandis} to conventional
superfluids.  We first note that for small $Q$, $K_{zz}(q,Q)\approx
K_{zz}(q,0) - \rho_s Q^2$ so that the maximum value of $Q$ is
$[\min(K_{zz}(q,0))/\rho_s]^{1/2}$.  One additional approximation is
required to obtain the same result from the Landau criterion; we must
assume that $K_{\varphi\varphi}(q,Q=0) \approx \rho_s q^2$ for $q$ up
to the value at which $\varepsilon^*(q)/q$ is minimized, actually a
reasonable approximation for quantum Hall superfluids.  Using $n = (2
\pi \ell^2)^{-1}$, the density of a full Landau level, we see that the
superfluid velocity is related to $Q$ by $v_s = \rho_s Q (2 \pi
\ell^2) /\hbar$.  The Landau criterion $v_s^2 <
\min[{\varepsilon^*(q)^2/q^2}]\approx (2 \pi \ell^2)^2
\min[{K_{zz}(q,0)}\rho_s]$ then leads to the same limit on $Q$ as the
$K_{zz}(q,Q) > 0$ stability criterion, that we use here to obtain the
QH superfluid phase boundary illustrated in
Figs.\ref{unisupercurrent},\ref{omega_Q}.

\section{Weak Links in Quantum Hall Superfluids ($\Delta_{SAS}=0$)}

\begin{figure}
\center
\epsfxsize 8.5cm \rotatebox{0}{\epsffile{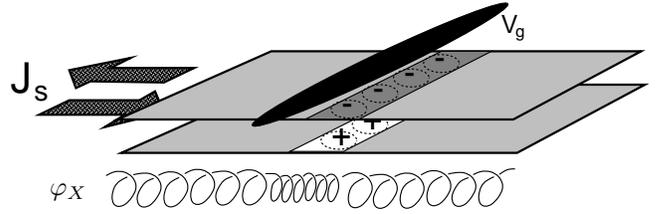}}
\begin{picture}(150,25)(0,0)
\put(-30,33){$\varphi_X$}
\end{picture}
\caption{Schematic illustration of a quantum Hall superfluid weak-link induced
  by a top gate.
  The pseudospin stiffness is reduced locally by a gate voltage that 
  displaces the bilayer system away from the balanced condition of equal
  density in the two layers.  The increased phase winding rate at the link,
  required by current conservation, is illustrated.}
\label{weak_link}
\end{figure}

One element of the physics of quantum Hall superfluids that might
potentially be interesting for future experimental study, is the
possibility\cite{wenzeeepl} of creating and tuning weak links using
gate voltages.  Weak links are locations in the superfluid at which
the local superfluid stiffness is reduced and phase changes are
therefore concentrated in a narrow region (link).  In quantum Hall
bilayer superfluids, one such weak-link geometry is illustrated in
Fig.\ref{weak_link}, where the local stiffness is reduced {\it in
  situ} by a gate voltage that locally moves the bilayer system away
from balanced condition of equal density in the two layers.

\begin{figure}
\center
\epsfxsize 6.0cm \rotatebox{-90}{\epsffile{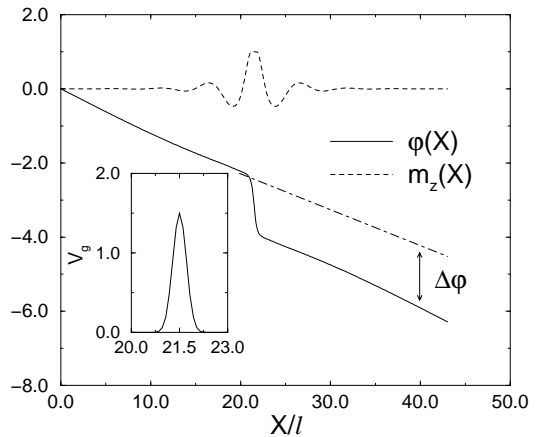}}
\vskip.5cm
\caption{$\varphi(X)$ and $m_z(X)$ are shown for a bilayer with
    $\Delta_{SAS}=0$ and $d=\ell$.  These results are for a weak link
    potential with a width equal to $\ell$ and a barrier height $1.5
    e^2/\epsilon \ell$.  $\Delta\varphi$ is measured with respect to
    its value at one edge of the periodic cell and the weak link is
    located at the center of the well.}
\label{weak_linkIV}
\end{figure}

In this section we discuss the supercurrent properties of quantum Hall
superfluids with such top gate-voltage induced weak links.  We
represent the electrostatic potential from a narrow gate separated
from the epitaxially grown bilayer system by a cap layer
(Fig.\ref{weak_link}), somewhat crudely, by an external potential that
has a narrow Gaussian variation in the $x$ direction, but is invariant
in the $y$ direction, as shown in the inset of Fig.\ref{weak_linkIV}.
The width of the Gaussian should be given approximately by the minimum
of the gate width and the thickness of the cap
layer.\cite{other_links}

In order to evaluate the Josephson-like relationship between the phase
change across a weak link and the superfluid current, we have solved
the mean-field equations for a periodic system system with one
weak-link and a $2 \pi$ phase change per period $a$.  In this way we
can vary the superfluid current by changing the period $a$.  A typical
solution is shown in Fig.\ref{weak_linkIV}.  

Since for vanishing interlayer tunneling, $\Delta_{SAS}=0$, considered
here the dc staggered supercurrent is spatially uniform, current
continuity demands $\partial \varphi/ \partial X$ to be inversely
related to the local superfluid density 
when the phase gradient is small. In particular, $\partial \varphi /
\partial X$ must be constant to an excellent approximation away from
the weak link.  We will refer to this constant as $Q$ below.  After
solving the mean-field equations for a series of $Q$ values, the
current density can be evaluated either by exploiting its relationship
to the $a$ dependence of the total energy or by combining the phase
winding rate $Q$ away from the weak link with the uniform system
$J_s(Q)$ expression.  The phase change across the weak link, $\Delta
\varphi$, is extracted by extrapolating the constant curve from one
side of the weak link to the other and finding the additional shift
that is present because of the weak link, as illustrated
Fig.\ref{weak_linkIV}.

\begin{figure}
\center
\epsfxsize 6.0cm \rotatebox{-90}{\epsffile{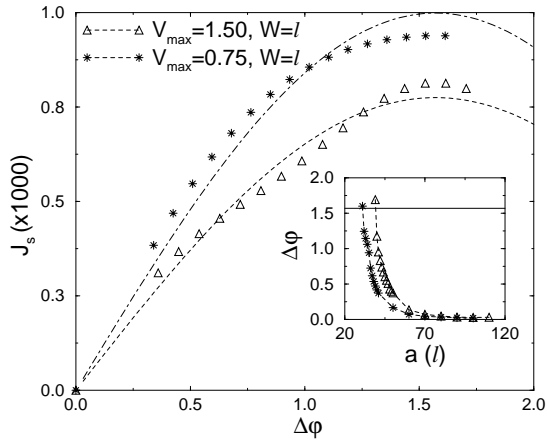}}
\vskip.5cm
\caption{Supercurrent density vs. weak-link phase slip angle
  $\Delta\varphi$ for weak links of width $\ell$ and heights
  $V_g^{max}=0.75$ and $V_g^{max}=1.50 e^2/\epsilon \ell$.  The
  maximum supercurrents in this case are approximately five times less
  than in the absence of a weak link.}
\label{Js_deltaphi}
\end{figure}

A typical result for $J_s$ {\it vs.} $\Delta \varphi$ obtained in this
way is summarized in Fig.\ref{Js_deltaphi}.  
\noindent For small currents (large $a$), $J_s$ increases linearly with $\Delta
\varphi$, as in a conventional Josephson junction.
Notice that $J_s$ reaches its maximum for $\Delta \varphi$
close (but not exactly equal) to $\pi/2$ as in the Ambegaokar-Baratoff
\cite{ambegaokar} theory of superconductors linked by a tunneling
Hamiltonian.  The microscopic physics of the coupling is quite
different in the quantum Hall superfluid, however.
For a weak link that is translationally
invariant along $y$\cite{other_links} (the geometry illustrated in
Fig.\ref{weak_link}) the guiding center $X$ remains a good quantum
number in the absence of disorder and consequently
the quasiparticles do
{\em not} tunnel across \cite{other_links}.  Instead, the coupling
between the uniform superfluids on opposite sides of such weak link is
due to Coulomb exchange interactions, both within and near to the weak link.
As in the case of a uniform superfluid, configurations of the order
parameter field that occur beyond the maximum of this function are not
stable, and the staggered critical current is given by the maxima of
these curves.

\begin{figure}
\center
\epsfxsize 6.0cm \rotatebox{-90}{\epsffile{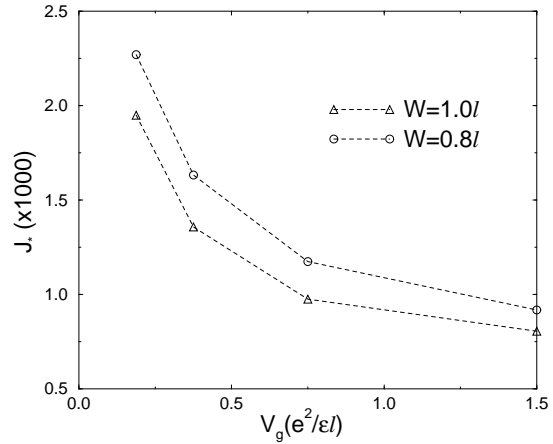}}
\vskip.5cm
\caption{Critical staggered supercurrent density dependence on the maximum
  of the bias potential $V_g^{max}$ responsible for the weak link,
  plotted for two width values.}
\label{Jsc_Vg}
\end{figure}

In Fig.\ref{Jsc_Vg} we plot critical currents as a function of the
strength of the bias potential that produces the weak link for two
different values of the width of the link.  The critical current
decreases rapidly with both the width of the barrier potential and its
strength.  Because sample inhomogeneities will undoubtedly introduce a
distribution of weak links into the superfluid bilayer, one conclusion
that follows immediately from these calculations is that the critical
current will likely sharply reduced by disorder in typical current
samples.

We close this section by noting that to obtain the weak-link critical
current $J_c$, we have relied solely on the vanishing
$K_{\varphi\varphi}$ criterion, and have ignored the $K_{zz}$
stability matrix. Although we have not checked this explicitly,
because a weak link substantially reduces the phase stiffness, leading
to a five-fold reduction in $J_*$, we believe it is this
$K_{\varphi\varphi}$ criterion that determines the weak-link critical
current, giving $J_c=J_*$.

\section{Role of Interband Hybridization in Quantum Hall Superfluids: 
The $\Delta_{SAS}\neq 0$ case}
\label{finiteDelta}

In the presence of interlayer tunneling $\Delta_{SAS}$, the superfluid
energy functional has soliton lattice extrema which break
translational symmetry but, as we comment later, still carry a finite
supercurrent.  It is not in general possible to address the
metastability of these solutions analytically, although a qualitative
understanding can be achieved using scaling arguments\cite{JordanLeo}.

\subsection{$K_{\varphi\varphi}$ instabilities of charge-balanced bilayers}

\begin{figure}
\center
\epsfxsize 6.0cm \rotatebox{-90}{\epsffile{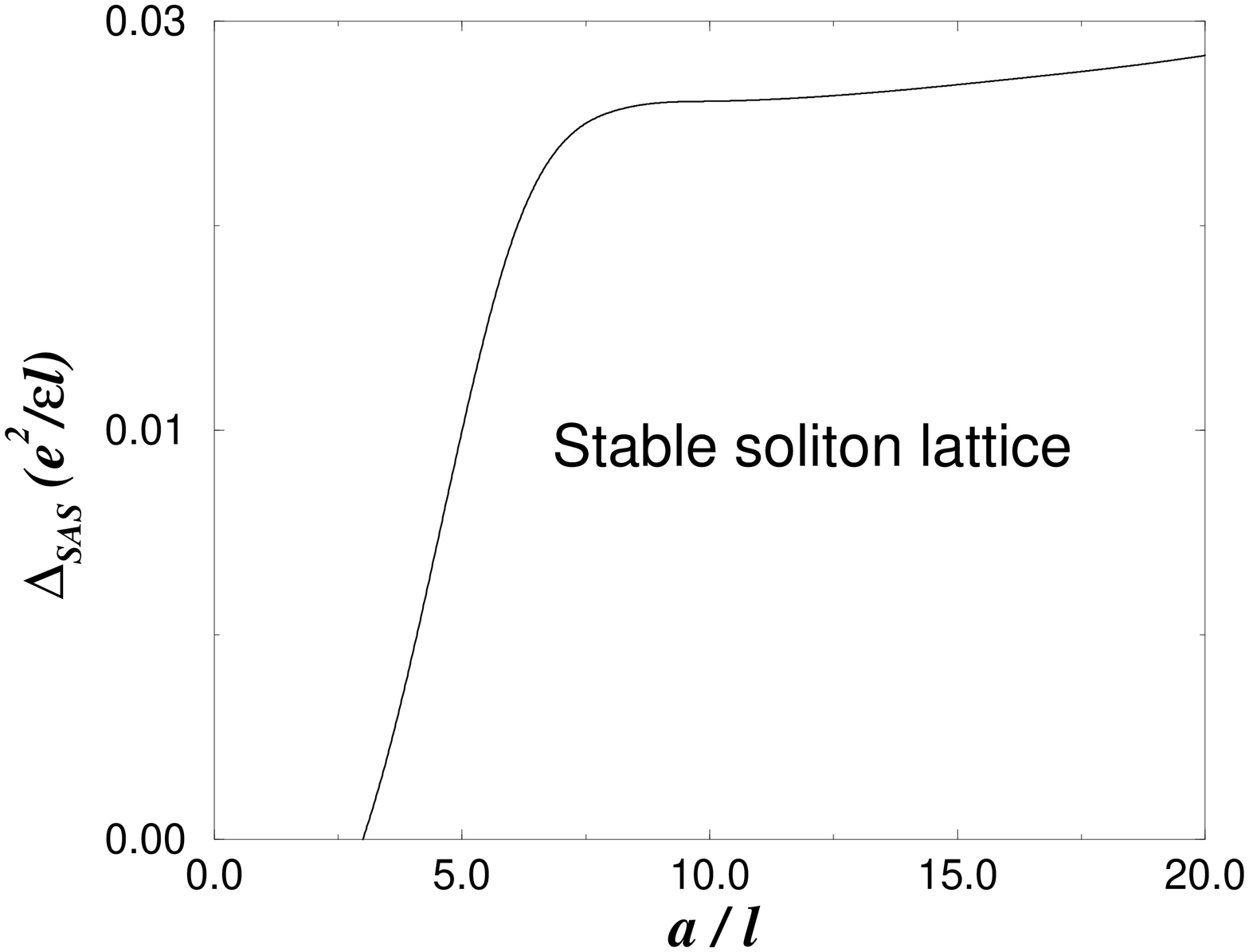}}
\vskip.5cm
\begin{picture}(150,25)(0,0)
\put(18,48) {\Large ${2\pi\over Q^*l}$}
\end{picture}
\caption{The stability region for soliton-lattice states, plotted as a
function of $\Delta_{SAS}$ and soliton spacing $a$ for $d = \ell$.
This phase diagram is constructed by solving the mean-field equations.
Metastability in the $\eta =0$ plane holds in the region labeled
Stable Soliton Lattice.}
\label{Fig7}
\end{figure}

To find members of this family of energy functional extrema, we first
solve the mean-field equations self-consistently to find $\varphi(X)$
solutions at $\eta(X) \equiv 0$.  We seek solutions with period $a$,
where $a$ is the distance between the soliton centers that are located
at the mid-points of our primitive cells.  For weak tunneling the
solitons are wide and are accurately\cite{Cote94} approximated by
solving the sine-Gordon equation that is obtained by minimizing the
gradient approximation to the $\eta \equiv 0$ energy functional.  It
follows that for dilute solitons at small $\Delta_{SAS}$, $\varphi(X)
\simeq 4 \tan^{-1}\exp[-(X-a/2)/\xi]$ modulo $2 \pi$ within each cell,
where $\xi=\ell\sqrt{4\pi\rho_s/\Delta_{SAS}}$ is the width of the
soliton.  We find numerically that as the density of solitons
increases (the unit cell period $a$ decreases) and the energy per
cross-sectional length increases, we eventually reach a situation at
$a=a_c$ for which the soliton lattice solution is no longer metastable
and the self-consistency procedure does not converge.  This point is
most conveniently identified numerically by reducing $a$ for a given
$\Delta_{SAS}$ until no mean-field solutions with finite phase winding
exist; we discuss a more systematic but numerically more cumbersome
approach below.  The result of this calculation is summarized in Fig.
\ref{Fig7} in which the regime of stable staggered-current-carrying
soliton lattice states is plotted as a function of $\Delta_{SAS}$ and
the soliton spacing $a$ for $d = \ell$.  For $\Delta_{SAS} \to 0$, the
soliton width diverges and the phase $\varphi(X)$ is described by a
plane-wave with wavevector $Q$.  The phase diagram that we calculate
here corresponds in this limit to the $K_{\varphi\varphi}$ stability
criterion and the maximum wavevector $Q_*$, agrees with that found
earlier (see Figs.\ref{Js0},\ref{unisupercurrent}) to within numerical
accuracy.

This phase boundary shape can be summarized by saying that there is
both a minimum distance between solitons and a maximum interlayer
tunneling amplitude at which metastable extrema of the energy
functional exist.  The physics behind this result is not difficult to
understand.  For weak tunneling, the phase gradient,
$\partial_X\varphi\approx 2\pi/\xi\propto\sqrt{\Delta_{SAS}}$ of an
isolated soliton ($a\gg \xi$) is below the critical value of $Q_*$
plotted as a dashed line in the inset of Fig.~\ref{unisupercurrent}.
In this regime the stability of the soliton state is therefore
predominantly controlled by the soliton spacing $a$, which, for a
sufficiently dense soliton lattice leads to phase gradient larger than
$Q_*$.  In contrast, for sufficiently large interlayer tunneling
$\Delta_{SAS}$ even an isolated soliton becomes unstable, when its
maximum phase gradient $\partial_X\varphi\approx
2\pi/\xi\propto\sqrt{\Delta_{SAS}}$ exceeds $Q_*$.  Consequently, in
the large $\Delta_{SAS}$, large soliton-spacing ($a$) regime, the phase
boundary is not sensitive to soliton spacing $a$, as reflected by the
approximately horizontal orientation illustrated in Fig.\ref{Fig7}. We
emphasize that the phase diagram constructed in this way does not
account for $K_{zz}$ instabilities.

The soliton lattice state carries a staggered supercurrent which can
be evaluated from Eq.~\ref{eq:current2} given mean-field-theory
numerical results for the dependence of the energy per unit length
perpendicular to the current direction $\varepsilon=8\rho_s^\perp/\xi$
on soliton density.  (The experimental significance of
Eq.~\ref{eq:current2} is discussed at greater length in Section VI.)
For each $\Delta_{SAS}$, the energy per unit length for low soliton
density ($a \to \infty$) is proportional to the soliton density and
the supercurrent therefore approaches a {\em minimum} value as $a \to
\infty$:
$J_{c1}=(4/\pi)\sqrt{\rho_s\Delta_{SAS}/\pi^3}$.\cite{Shevchenko,JordanLeo}
Evidently $J_{c1}$ decreases by decreasing the interlayer tunneling,
and it vanishes like $\Delta_{SAS}^{1/2}$ as $\Delta_{SAS} \rightarrow
0$.
\begin{figure}
\center
\epsfxsize 6.0cm \rotatebox{-90}{\epsffile{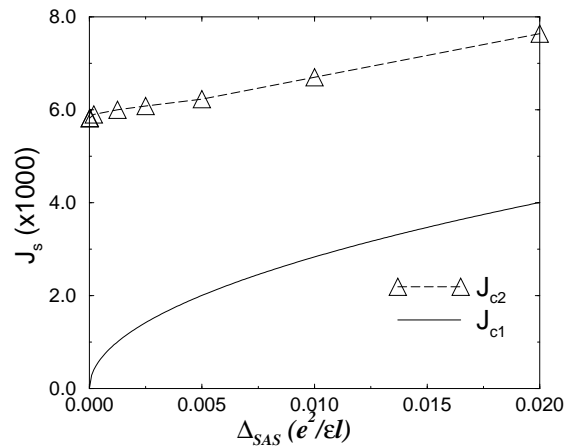}}
\vskip.5cm
\caption{Maximum and minimum for the persistent current density of 
  soliton lattice states in quantum Hall ferromagnets {\em vs.}
  $\Delta_{SAS}$.  $J_{c1} \propto \sqrt{\Delta_{SAS}}$ for
  $\Delta_{SAS} \to 0$.}
\label{Jsmax_Delta}
\end{figure}
At large staggered current we also find a {\em maximum} sustainable
supercurrent density $J_{c2}$, that corresponds to a large soliton
density.  In this dense regime, defined by $\xi/a \gg 1$ solitons
overlap and current is nearly uniform, leading to a highly oscillating
tunneling energy that therefore averages away to zero. Consequently,
we expect that $J_{c2}$ to be well-approximated by $J_*$ arising from
our earlier criteria of $K_{\varphi\varphi}=0$ for $\Delta_{SAS}=0$.
Consistent with this, we find that $J_{c2}$, shown in
Fig.\ref{Jsmax_Delta}, is not very sensitive to the tunneling matrix
element, increasing only slowly with $\Delta_{SAS}$.

The experimental meaning of $J_{c1}$ is unclear since one
might expect to find solutions to generalized sine-Gordon equations
for which the spatially averaged staggered supercurrent in a finite
length systems has an arbitrarily small value, for example by having
only a partial soliton across the entire system.  
A staggered current imposed at the edge of the system would fix 
$\partial_X\phi_X| \propto J_s$ only near the boundaries of the
bilayer.  Provided that these boundary conditions are physically 
realistic, a question that would require 
further microscopic analysis to verify, extrema of the functional
exist for all current values at the edge, and for all spatially
averaged staggered currents.  We discuss this point again in Section VI.

\subsection{$K_{zz}$ and $K_{\varphi\varphi}$ staggered current 
  instabilities}
  
\begin{figure} \center \epsfxsize 6.0cm
  \rotatebox{-90}{\epsffile{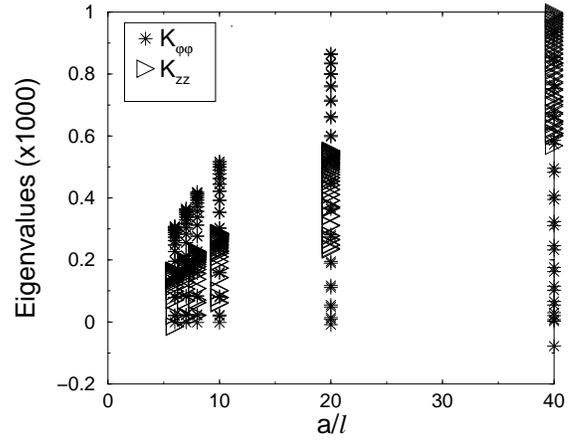}}
  \vskip.5cm
\caption{Stability matrix eigenvalues for $N=400$ and $\Delta_{SAS}=0$
  as a function of phase period $a$.  The pairing wavevector
  $Q=2\pi/a$.}
\label{Kzz_eigen_Delta_SAS_0_N_400_a}
\end{figure}

\begin{figure}
\center
\epsfxsize 6.0cm \rotatebox{-90}{\epsffile{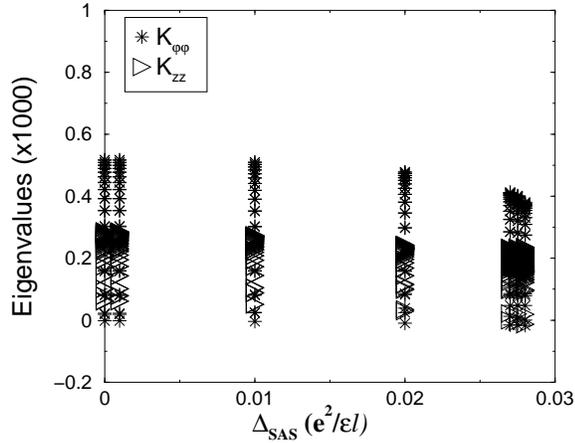}}
\vskip.5cm
\caption{Stability matrix eigenvalues for $N=400$ and $a=10 \ell$ as a 
  function of $\Delta_{SAS}$.}
\label{Kzz_eigen_Delta_SAS_N_400_a10}
\end{figure}

\begin{figure}
\center
\epsfxsize 6.0cm \rotatebox{-90}{\epsffile{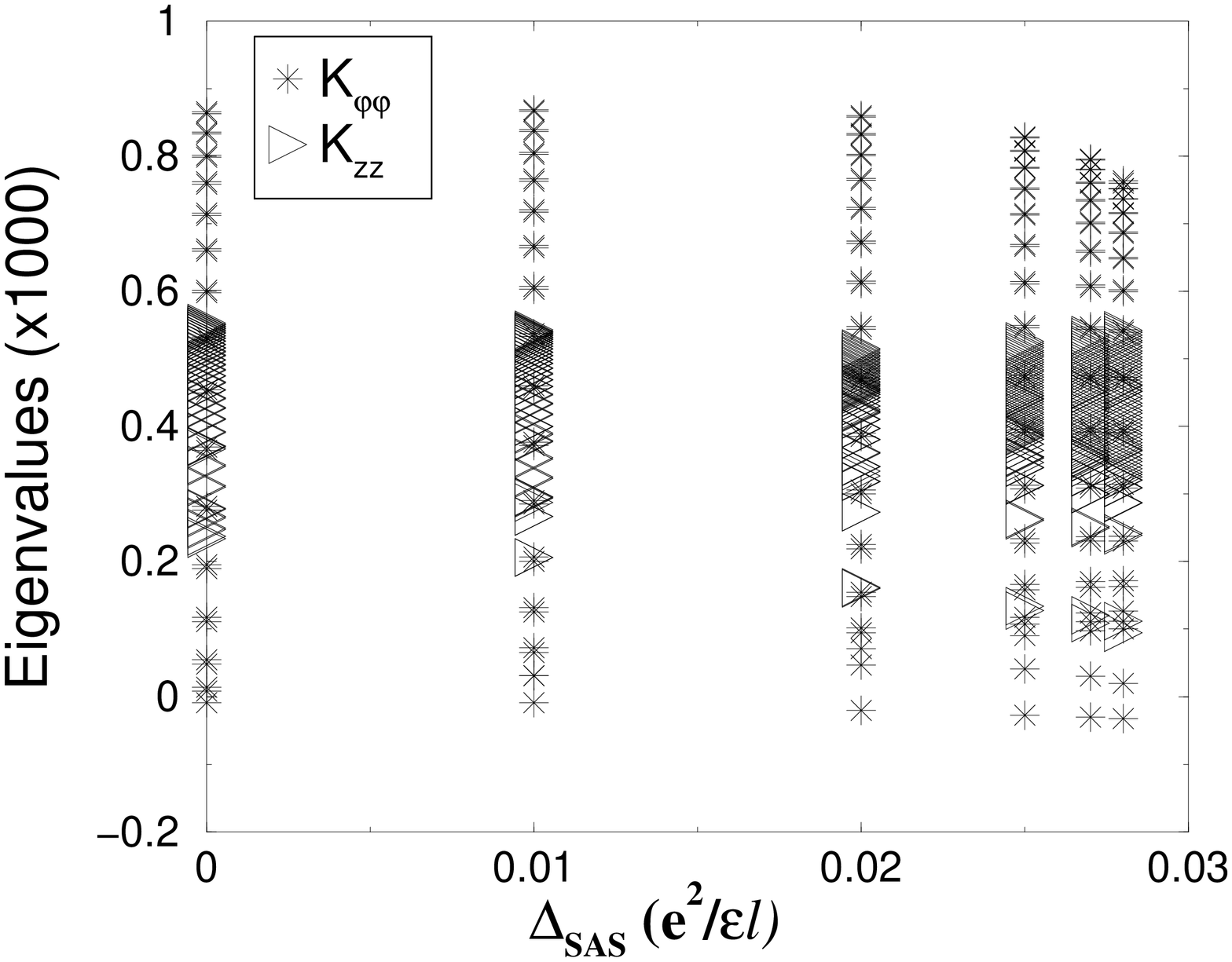}}
\vskip.5cm
\caption{Stability matrix eigenvalues for $N=400$ and $a=20\ell$ as a 
  function of $\Delta_{SAS}$.}
\label{Kzz_eigen_Delta_SAS_N_400_a20}
\end{figure}

\begin{figure}
\center
\epsfxsize 6.0cm \rotatebox{-90}{\epsffile{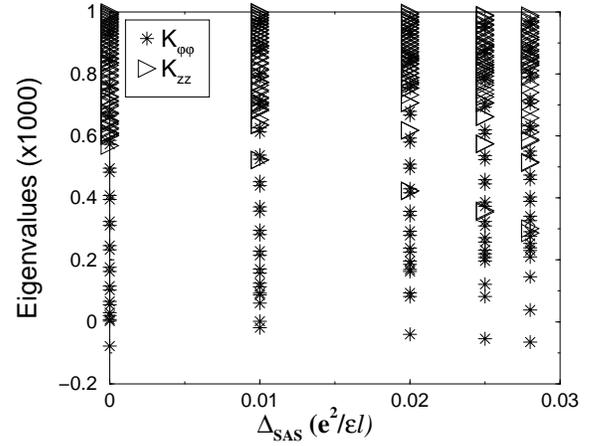}}
\vskip.5cm
\caption{Stability matrix eigenvalues for $N=400$, and $a=40\ell$ as a 
  function of $\Delta_{SAS}$.}
\label{Kzz_eigen_Delta_SAS_N_400_a40}
\end{figure}

A more systematic method to determine whether or not a solution of the
mean-field equations is metastable is to calculate the eigenvalues of
{\em both} matrices $K_{zz}(X,X')$, and $K_{\varphi\varphi}(X,X')$,
evaluated at the corresponding energy functional extremum.  We have
made some progress with this approach, although we have found it to be
(unsurprisingly) numerically cumbersome.  To make the calculation
feasible, we take the length $L_y$ of the system in the $y$ direction
to be finite, thereby discretizing the set of allowed guiding centers
along the $\hat x$-axis.  For $N$ guiding centers per soliton,
$K_{zz}(X,X')$ and $K_{\varphi\varphi}(X,X')$ are $N \times N$
matrices.  In this study we increase $N$ as far as is practical, then
attempt to extrapolate to $N=\infty$.  Our numerical results are
summarized in Figs.
(\ref{Kzz_eigen_Delta_SAS_0_N_400_a}-\ref{Kzz_eigen_Delta_SAS_N_400_a40})
for a bilayer electron system with $d=\ell$.

In discussing these results, it is instructive to start with
$\Delta_{SAS}=0$, since we can compare with the results obtained
analytically in previous sections by Fourier transformation.  In Fig.
\ref{Kzz_eigen_Delta_SAS_0_N_400_a}, we present spectrum of $K_{zz}$,
and $K_{\varphi\varphi}$ evaluated at $\Delta_{SAS}=0$ and $d = \ell$.
Starting from a large soliton spacing $a$ (low staggered current), we
see that the lowest eigenvalue of $K_{zz}$ decreases with decreasing
$a$, finally vanishing for $a \sim 5.2 \ell$, consistent with the
value of $Q$ at which the $K_{zz}$ instability occurs in
Fig.\ref{Kzz_a20}. Because of the global phase invariance at
$\Delta_{SAS}=0$, there should always be one zero eigenvalue of
$K_{\varphi\varphi}$ at any value of $a$, an expectation that we
verify within numerical accuracy.  It appears from
Fig.\ref{Kzz_eigen_Delta_SAS_0_N_400_a}, that the finite values of
$L_y$ that we use for the case of inter-soliton distance $a = 40 \ell$
($L_y = 20 \pi \ell$) are sufficiently short that we introduce
artificial instabilities that do not occur for $L_y \to \infty$.  For
smaller values of $a$, we are able to consider larger values of $L_y$
without making our matrices inconveniently large.  These calculations
were performed with $L_y = 800 \pi \ell^2 / a$.  For $a = 40 \ell$ we
have verified that the negative eigenvalue becomes smaller in
magnitude, decreasing toward zero as $L_y$ is increased.  Consistent
with the analytical results, we find a large number of higher
eigenvalues of $K_{\varphi\varphi}$ approach zero as $a$ approaches
$a^* \simeq 3.9 \ell$, the value of $a=2\pi/Q$ at which the phase
instability occurs for $\Delta_{SAS}=0$.  The smallest magnitude
non-zero eigenvalue of $K_{\varphi\varphi}$ becomes negative at
$a^*=3.9\ell$, in reasonable agreement with the location of the
supercurrent peak in Fig.\ref{Js0}.

In Fig.~\ref{Kzz_eigen_Delta_SAS_N_400_a10} we show how the stability
matrix spectra depend on $\Delta_{SAS}$.  We find that for a given
value of $a$, increasing $\Delta_{SAS}$ always leads to an
instability.  For $K_{\varphi,\varphi}$ this finding is consistent
with the mean-field calculation results discussed above.  The present
results indicate that $K_{zz}$ becomes unstable near the same value of
$\Delta_{SAS}$ as $K_{\varphi,\varphi}$; because of the finite values
we must use for $L_y$ we have not been able to determine which
instability occurs first with a high degree of certainty.
Figs.\ref{Kzz_eigen_Delta_SAS_N_400_a10}-\ref{Kzz_eigen_Delta_SAS_N_400_a40},
indicate that negative eigenvalues in $K_{zz}$ do appear quite
generally when the tunneling energy increases.  The instabilities seem
to appear first in $K_{zz}$ for small $a$ and first in
$K_{\varphi,\varphi}$ for large $a$.  At $a=10\ell$ the smallest
eigenvalue of $K_{zz}$ becomes negative for $\Delta_{SAS} \approx
0.028 e^2/(\epsilon\ell)$, near where our the mean-field theory $L_y
\to \infty$ results indicate the first $K_{\varphi,\varphi}$
instability.  Interestingly, these instabilities at finite
$\Delta_{SAS}$ have no counterpart in the sine-Gordon theory of the
soliton lattice state.  They can be understood as being analogous to
the maximum currents and pairing wavevectors $Q$ that we find in the
translationally invariant case.  As $\Delta_{SAS}$ increases the
maximum rate of phase winding at the center of a soliton, $Q_{eff} =
\partial \varphi / \partial X$, increases.  For an isolated soliton in
the sine-Gordon model $Q_{eff} \approx
2\pi/\xi\propto\sqrt{\Delta_{SAS}}$.  Thus a maximum tunneling
amplitude $\Delta_{SAS} \sim 0.3 e^2/\epsilon \ell$ is not unexpected
given our results for the homogeneous case.

\section{Discussion and Conclusions} 

In this paper we have presented a mean-field theory of staggered
supercurrent carrying states in quantum Hall bilayer superfluids.  Our
analysis was based primarily on the behavior of two stiffnesses
$K_{zz}$ and $K_{\varphi\varphi}$ that characterize respectively the
energy cost of small interlayer charge and interlayer phase
fluctuations.  By computing these quantities, we have been able to
address the physics that controls the dependence of critical staggered
current on interlayer tunneling amplitude, layer
spacing, and other tunable parameters that characterize bilayer
systems.  We also studied the characteristics of weak links in
quantum Hall superfluids that are created by a gate voltage.  Unlike
the field-theoretic approach\cite{JordanLeo}, the microscopic analysis
here does not rely on a small wavevector expansion, and therefore has
the virtue of accounting for the full wavevector dependence of the
stability matrices.  This is especially important because the
wavevector expansion of these matrices is not analytic which
indirectly causes these quantities to have minima at finite
wavevectors.  The staggered critical supercurrent predictions
presented in this paper are consequently more realistic
than those derived from the field-theoretic approach\cite{JordanLeo}.

\begin{figure}
\center
\epsfxsize 8.5cm \rotatebox{0}{\epsffile{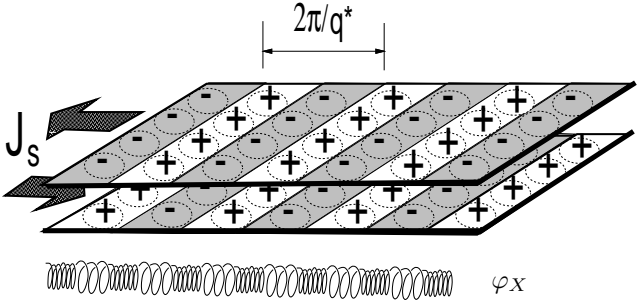}}
\begin{picture}(150,25)(0,0)
\put(140,33){$\varphi_X$}
\end{picture}
\caption{Schematic representation of the staggered current-induced QH
  supersolid state for which periodic interlayer charge modulation and
  QH superfluidity coexist. The periodic rate of phase winding is also
  illustrated.}
\label{supersolid}
\end{figure}

There are a number of limitations to the approach we have followed.
Our analysis is confined to {\em harmonic} instabilities and will
therefore miss any that are induced through a nonlinear mode, nucleation of
vortices for example.  In addition, our harmonic analysis does not
unambiguously determine the state to which the system evolves when the
instability point is reached. Based on experience with ordinary
superfluids and superconductors, a natural guess is that once one of
the stiffnesses vanishes, the current-carrying superfluid state of the
quantum Hall bilayer is unstable to a resistive (normal) state in which  
staggered current leads to a staggered voltage. We believe that this
scenario applies when $K_{\varphi\varphi}$ controls the instability,
and when the instability in $K_{zz}(q)$ takes place at a large
wavevector corresponding to a length scale of order or smaller than
the magnetic length $\ell$.  The instability process has that
character for the large interlayer spacing values, $d\gtrsim \ell$,
relevant to current experiments.  However, in the opposite limit, $d
<< \ell$, it is likely (based on similarity with our earlier work on
the interlayer-charge imbalance instability driven by an in-plane
magnetic field\cite{LRcanting,RaminLeo}) that vanishing $K_{zz}(q^*)$
instead signals an interlayer charge imbalance quantum
(nonequilibrium) transition to a distinct staggered current carrying
QH superfluid state.  As with the equilibrium canting
transition\cite{LRcanting,RaminLeo} at small $d/\ell$, the instability
signaled by vanishing of $K_{zz}(q^*)$ occurs at a small but finite
wavevector $q^*$ (with $q^*\ell << 1$), and therefore corresponds to a
development of a dipole-stripe state akin to a unidirectional charge
density state with period $2\pi/q^*$, as illustrated in
Fig.\ref{supersolid}.  Since, in addition to staggered gauge symmetry,
such a state also spontaneous breaks translational invariance along
the staggered current direction, it corresponds to a {\em quantum Hall
  supersolid} with coexisting superfluid, solid, and quantum Hall
orders.
\begin{figure}
\center
\epsfxsize 9.0cm \rotatebox{0}{\epsffile{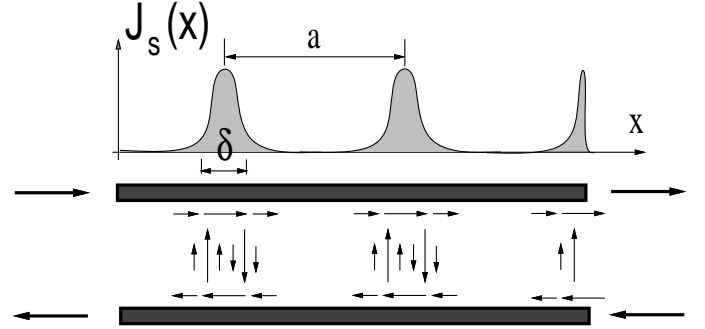}}
\vspace{0.5cm}
\caption{Schematic QH bilayer in the presence of finite
  single-particle interlayer tunneling $\Delta_{SAS}$.  When there is
  collective tunneling between the layers, the staggered supercurrent
  must be spatially dependent in order to conserve the difference in
  charge between the layers.  The spatially averaged staggered
  supercurrent can be arbitrarily small in a finite systems because of
  edge effects.  The soliton width in this illustration
  $\delta=\ell\sqrt{\rho_s/\Delta_{SAS}}$, while the period $a$ is the
  distance between soliton centers.  This schematic plot shows a
  partial soliton on the right that can make an arbitrarily small
  contribution to the spatially averaged staggered supercurrent.}
\label{soliton}
\end{figure}

%
Another feature of our work is the finding that the staggered super
current evaluated for soliton-lattice states using
Eq.~\ref{eq:current2} has a minimum value $J_{c1}$ as well as a
maximum value.  This expression for the staggered current can be
motivated in several different ways, but its connection to experiment
is not always direct, as we now explain.  The general expression for
the staggered current can be derived by introducing a spatially
constant staggered vector potential $A_{\uparrow} = - A_{\downarrow} =
A/2 $ and noting that the interaction energy is then a function of the
gauge invariant inter-layer phase difference $\varphi_X + e X A/ \hbar
c$.  It follows that the staggered current is given by
\end{multicols}

\widetext
\begin{eqnarray} 
J_s &=& \frac{1}{\hbar L_x L_y} \frac{\partial E}{\partial
  A_{\uparrow}} \nonumber \\
&=& \frac{1}{2 \hbar L_x L_y^2} \sum_{X,X'} F_D(X-X') \sin\theta(X) 
\sin\theta(X') \sin [\varphi(X)-\varphi(X')+e(X-X')A/\hbar c] (X-X').
\label{eq:currentgeneral}
\end{eqnarray}
\noindent

\begin{multicols}{2}
\columnwidth3.4in
\narrowtext
\noindent
The vector potential $A$ has no physical meaning if $\Delta_{SAS}=0$.
For $\Delta_{SAS} \ne 0$, however, an in-plane field is represented by
$A = B_{\parallel} d$ and in equilibrium Eq.~\ref{eq:currentgeneral}
describes the persistent currents responsible for orbital
diamagnetism.  In the absence of an in-plane field $A=0$, and
$\varphi(X)$ is determined by minimizing Eq.~\ref{HF}.  When the
minimization procedure is carried out with periodic boundary
conditions, as in our numerical calculations, changing $A$ is
equivalent to changing the period over which $\varphi$ is required to
change by $2 \pi$, and Eq.~\ref{eq:current2} follows.  For realistic
systems, however, edge effects can have a large effect on the current,
especially when individual solitons, which carry circulating currents
in their cores, are well separated.

%
$J_{c1}$ is analogous to the lower critical magnetic field
$H_{c1}$ in type-II superconductors.  Only for $J_s > J_{c1}$ does it
pay for the system to give up some of the tunneling kinetic energy and
allow a soliton that carries this minimum staggered current $J_{c1}$
into the system.  $J_{c1}$ is closely related to the
critical in-plane magnetic field that induces the
commensurate-incommensurate transitions in quantum Hall
bilayers.\cite{yangmoon,LRcanting}

Consider, for example, the limit of slowly varying phases for which
the local staggered current density can be shown to be\cite{yangmoon}
$2 \rho (\partial \varphi/\partial x)$, in agreement with
Eq.~\ref{eq:currentgeneral}.  In this case the spatially averaged
staggered current is proportioanl to the number of solitons in the
system, which can clearly be arbitrarily small.  This conclusion
should be contrasted with that reached by Shevchenko\cite{Shevchenko},
working with a closely related long-wavelength model that applies to
the case of of electron-hole bilayers at zero magnetic field.
(Excitonic superfluidity has not yet been conclusively established in
experiment for this case.)  Shevchenko's conclusion that $J_{c1}$
would represent an experimental minimum staggered supercurrent was
based on the argument that finite current configurations could be
studied by adding a soliton chemical potential term [more precisely a
``chemical potential'' for interlayer phase twist] $J_s 2\rho_s
\partial_X\varphi/ \hbar $ to the bilayer energy density functional.
This term is chosen so that the modified energy functional has its
absolute minimum when Eq.~\ref{eq:current2} for the current density is
satisfied, and is modelled after related approximations that are
common\cite{cbjj} in the description of current-biased Josephson
junctions.  In this approach, because for finite tunneling
$\Delta_{SAS}$ the induced phase twist costs interlayer kinetic energy
(tunneling), a minimum staggered current $J_{c1}$ is required for
phase winding to occur.  Below this critical value, $J_s< J_{c1}$
(analogous to a Meissner state of type-II superconductor, where vortex
density $\propto B$ vanishes) no solitons are induced, and phase
$\varphi_X$ remains spatially uniform and pinned at $0$ (except for
the boundaries). The consequent vanishing of the interlayer tunneling
current leads to a conserved staggered current $J_s(x)$, that is
therefore spatially uniform.
In our view, however, this thermodynamic description of finite current
states does not apply to electron-hole bilayer systems with excitonic
superfluidity, either at zero field or in the quantum Hall regime.
The key difference between quantum Hall bilayers and Josephson
junctions is the absence of a superconducting current-bias environment
which imposes a definite value of the staggered curent at every point
in space.  In the bilayer quantum Hall case, current-carrying states
are never equilibrium states, but they can be metastable.  External
contacts can in principle, impose a given value of the staggered
current at every point on the {\em edge} of the system, assuming that
non-collective current carrying excitations can be completely
discounted.  However, the staggered current density at interior points
must be determined by finding local minima of the energy functional
that are consistent with edge boundary conditions.  In general when
$\Delta_{SAS}$ is non-zero, many such solutions exist.  The critical
current is naturally defined as the metastable configuration with the
largest staggered supercurrent, even for geometries that are more
general than those considered here.

It is difficult for several reasons to use our microscopic HFA
approach to calculate the activation energies for the saddle point
extrema that separate local minima of the HF energy with different
numbers of phase slips.  These activation energies will control the
thermal phase-slip nucleation rate and the associated staggered
voltage at any finite temperature.  One problem encountered in seeking
these solutions is that standard simple iterative approaches can be
used to find solutions of the mean-field equations that correspond to
local minima, whereas some more subtle and clever technique would have
to be used to coax the system iteratively toward a saddle-point
solution.  A second problem is that for the wide Hall bars of interest
to us here, the saddle point solution likely consists of nucleating a
vortex and moving it to the middle of the Hall bar; these solutions
have polar and azimuthal angle fields that depend on both spatial
coordinates, making the problem two-dimensional.

\begin{figure}
  \center \epsfxsize 8.0cm \rotatebox{0}{\epsffile{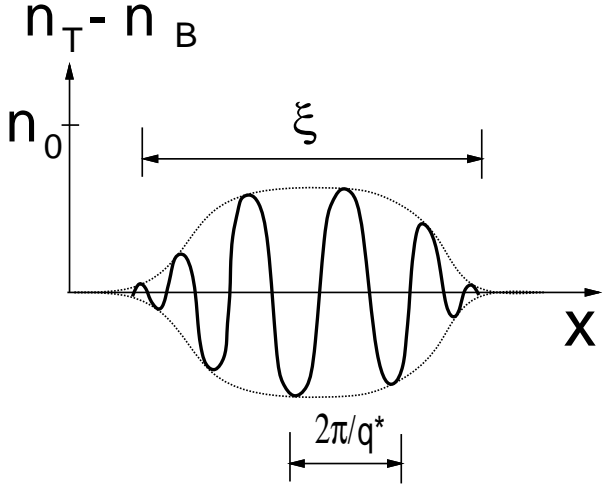}}
\vspace{0.5cm}
\caption{Schematic illustration of the saddle-point solution,
  $m_z^{sp}(x)$ for $\xi_0 q_*\rightarrow 1^-$, characterized by
  envelope width $\xi$ and wavevector $q_*$.  The free-energy
  associated with this $m_z^{sp}(x)$ controls the energy barriers for
  thermal activation of phase slips and dissipation in the staggered
  current channel when the Hall bar is narrow.}
\label{mzSP}
\end{figure}

We can, however, make some progress toward understanding staggered
channel dissipation by combining our HFA results and a scaling
analysis supported by exact field-theoretic calculations valid for a
short-range interaction model\cite{JordanLeo}.  For example, consider
the case of thermally-activated staggered resistance controlled by a
saddle-point phase-slip solution $m_z^{sp}(x)$, similar to the one
illustrated in Fig.\ref{mzSP}.  We expect that the limiting saddle
point will have this character for narrow Hall bars.  The phase slip
is characterized by a single length scale
\begin{equation}
\xi={\xi_0\over\sqrt{1-\xi_0^2 q_*^2}},\label{xi}
\end{equation}
which reduces to the width
$\xi_0=\sqrt{\rho_s/K_{zz}(0)}$\cite{JordanLeo} of saddle-point
solution in the limit of short-range interactions or small $\xi_0
q_*$, valid for $d/\ell\ll 1$; $q_*$ is the wavevector at which, in
the presence of long-range Coulomb interaction $K_{zz}(q)$ is minimum.
In this short-range limit, we expect the saddle-point solution
$m_z^{sp}(x)$ and the associated energy barrier to be well described
by the field-theoretic prediction\cite{JordanLeo}. 

In the opposite limit, as the width $\xi_0$ of the soliton nucleation
site approaches (from below) the period $2\pi/q_*$ at which the
interlayer charge-imbalance $m_z=n_T-n_B$ instability takes place, we
expect a saddle-point solution to be qualitatively modified by
long-range Coulomb interactions. In this regime we predict an
oscillatory $m_z^{sp}(x)$, with wavevector $q_*$ and an envelope of
width $\xi$, as illustrated in Fig.\ref{mzSP}. The associated energy
barrier in the limit of vanishing staggered current is given by
\begin{equation}
E_B = \epsilon_B^0(1-\xi_0^2 q_*^2)^{1/2}L_y,\label{E_B}
\end{equation}
where $\epsilon_B^0$ the energy barrier per unit of soliton length in
the limit of short-range interactions\cite{JordanLeo}. 

This analysis has been limited so far to one-dimensional
instabilities. The energy barrier associated with such a
one-dimensional soliton grows with the width $L_y$ of the quantum Hall
bar. In contrast, for vanishing tunneling $\Delta_{SAS}=0$, barriers
for {\em point}-vortex nucleation are proportional to $\rho_s$, up to
corrections logarithmic in $L_y$.  Consequently, for sufficiently wide
quantum Hall bars, $L_y > \xi$\cite{JordanLeo}, the staggered
resistivity will be determined by point-vortex nucleation and
unbinding.  Indeed standard arguments guarantee that some dissipation
will always be present at finite temperatures and voltages, as we
discuss below.

Our discussion has so far ignored quantum and thermal fluctuations and
quenched disorder, present in real systems. For layer separations
$\sim 10 \%$ below the critical value
$d_c$\cite{joglekargrpa,schliemann}, quantum fluctuations in the
absence of quenched disorder can be treated
perturbatively\cite{joglekargrpa} and will yield only small
corrections to our results. In contrast, thermal fluctuations and
especially in combination with disorder can modify some of our results
qualitatively.  Even in the absence of quenched disorder, for
vanishing tunneling, thermal fluctuations lead to a finite, activated
dissipation (i.e., staggered voltage) at any finite staggered current,
and therefore as in the case of superconductors, preclude an
unambiguous definition of a critical current. In this case a
superfluid state is distinguished from a dissipative one by a
vanishing {\em linear} resistivity. Extending standard
arguments\cite{Ambegaokar} to our system for $\Delta_{SAS}=0$ limit,
we predict a nonlinear power-law staggered I-V characteristic $E_s\sim
J_s^\alpha$, $\alpha(T)\geq 3$ to characterize the quantum Hall
staggered superfluid transport.  Despite finite dissipation even below
mean-field critical staggered current $J_c$ computed here, at low
temperatures we nevertheless expect a strong crossover in the I-V at
$J_c$, with the staggered voltage drastically dropping for current
below $J_c$ and dissipation controlled by thermally activated vortex
nucleation and vortex transport over barriers.
\begin{figure}
  \center \epsfxsize 8.0cm \rotatebox{0}{\epsffile{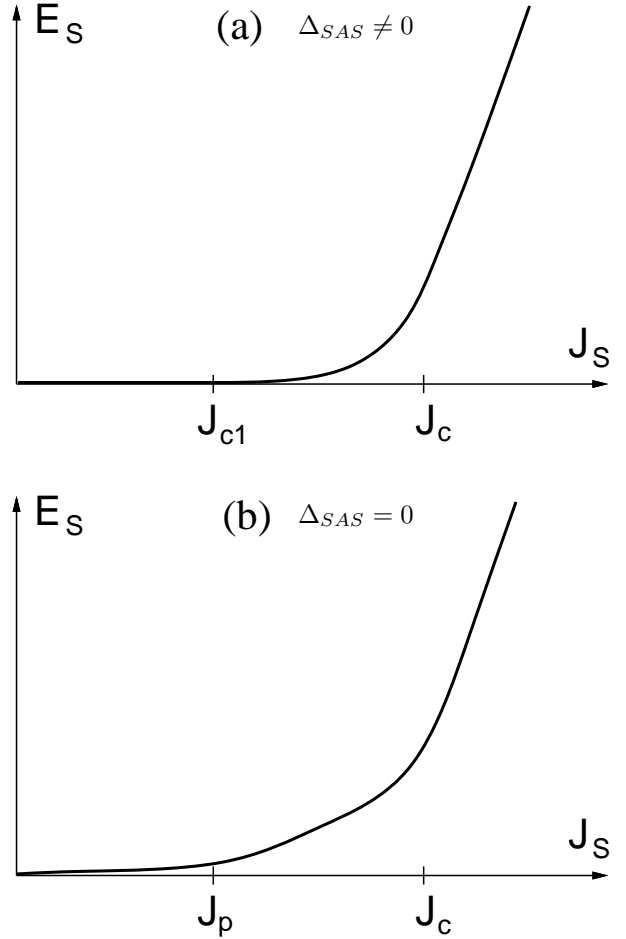}}
\begin{picture}(150,25)(0,0)
\put(70,372){$\Delta_{SAS}\neq0$}
\put(70,187){$\Delta_{SAS}=0$}
\end{picture}
\caption{Low temperature and weak disorder staggered current-voltage
  (I-V) characteristics for QH bilayers for (a) finite and (b)
  vanishing interlayer tunneling $\Delta_{SAS}$.}
\label{IV}
\end{figure}

For finite interlayer tunneling, $\Delta_{SAS}\neq 0$, the ionization
energy necessary to separate vortex pairs grows linearly with
separation $R$ as $(\rho_s\Delta_{SAS})^{1/2}R/\ell$.  Consequently,
in thermodynamic limit (wide bilayers) we should expect a true
staggered critical current
$J_{c1}=(1/\hbar\ell)(\rho_s\Delta_{SAS})^{1/2}$, below which
staggered transport is dissipationless, $E(J_s<J_{c1})=0$, even at
finite temperature, and for $J_s>J_{c1}$,
$E(J)\propto|J_s-J_{c1}|^{\alpha(T)}$\cite{JordanLeo}. The
corresponding I-V is illustrated in Fig.\ref{IV}a.

Quenched disorder is another important ingredient that we have not
considered in detail here.  As we have already emphasized, our
analysis of the weak-link strongly suggests that by creating weak
links, disorder will suppress critical currents.  Short-range disorder
potential fluctuations will tend to nucleate vortices, a fact that can
be understood in terms
of the short-range-correlated random effective vector
potential\cite{stern99,tuntheory} that it gives rise to.  For
$\Delta_{SAS}\rightarrow0$, the random vector potential model is known
to have a stable low temperature quasi-long-range ordered phase with
no unbound vortices that appears below a critical strength of
disorder.\cite{nattermann} (Microscopic
calculations\cite{joglekargrpa} that treat disorder using a
self-consistent Born approximation, lead to the same conclusion.)
Beyond the presumed disorder threshold, the system is in the
gauge-glass regime, which in two dimensions is unstable to vortices
that are mobile at finite temperature and only localize as
$T\rightarrow0$. Hence in the weak disorder limit, when vortices are
dilute, $\delta \ll L_L$ (where $L_L$ is the Larkin length), the
mean-field critical current found in this paper should be clearly
observable. In fact for weak disorder we expect to see three regimes
of dissipation via motion of disorder-induced vortices: (i) slow,
thermally activated creep for $J_s < J_p$, (ii) vortex flow, for $J_p
< J_s < J_c$ (where $J_c$ is the mean-field critical current of a
clean system derived here) , and (iii) normal state dissipation for
$J_c<J_s$, when the system is past its point of metastability. The
corresponding I-V is schematically illustrated in Fig.\ref{IV}b.

How then might the critical currents calculated here, which are based
on a Landau-like metastability requirement for the staggered-current
carrying state, appear experimentally?  The best possibility appears
to be by looking for features in the non-linear I-V relationships seen
in interlayer quantum Hall drag experiments.\cite{jpe3} So far these
experiments seem to show a linear relationship between the
longitudinal staggered voltage and the staggered current, with a
thermally activated dissipation process.  There is no hint at present
that the staggered electric field initially grows sublinearly with
current, suggesting that $J_p$ might be driven to zero by disorder.
It will be very interesting to see whether or not a relatively sharp
crossover can be observed experimentally at low temperatures at a
current density that can be associated with $J_c$.  The crossover
would be from a regime in which the staggered current is largely
collective, but not completely dissipationless due to the motion of
vortices nucleated thermally and by disorder, to a regime in which
there is essentially no collective staggered current.  If our
estimates are qualitatively reliable, the scale of $J_c$ is normally
larger than those used in typical experiments but vanishes as
$(d_c-d)^{1/2}$ as the phase boundary for the spontaneous coherence
state is approached from the below.  For this reason, we believe that
nonlinear transport experiments in the {\em drag} geometry, current in
one layer and voltages measured in both layers, as a function of
$d_c-d$ will be helpful in sorting out the complex physics present in
current samples.\cite{jpe3}

\section{Acknowledgements}

We gratefully acknowledge helpful interactions with Leon Balents,
Anton Burkov, Herb Fertig, Steve Girvin, Bert Halperin, Yogesh
Joglekar, Jordan Kyriakidis, Enrico Rossi, John Schliemann, and Ady
Stern.  This work was supported by the Welch Foundation, by the
National Research Council under grants DMR-0115947 and NSF MRSEC
DMR-0080054, and by the Sloan and Packard Foundations.


\end{multicols}
\end{document}